\documentclass[%
 11pt,
 tightenlines,
nofootinbib,
 amsmath,amssymb,
 aps,
 pra
]{revtex4-2}

\usepackage{graphicx}% Include figure files
\usepackage{xcolor}
\usepackage{amsthm}
\usepackage[margin=1in]{geometry}
\usepackage[pdftex,colorlinks=true,linkcolor=blue,citecolor=blue,urlcolor=black]{hyperref}
\usepackage{comment}

\newcommand{\R}{\mathbb{R}}

\newcommand{\Z}{\mathbb{Z}}

\newcommand{\E}{\mathbb{E}}

\newcommand{\ket}[1]{| #1 \rangle}

\newcommand{\braket}[1]{\langle #1 \rangle}

\DeclareMathOperator{\poly}{poly}

\newcommand{\be}{\begin{equation}}
\newcommand{\ee}{\end{equation}}
\newcommand{\bea}{\begin{eqnarray}}
\newcommand{\eea}{\end{eqnarray}}
\newcommand{\bes}{\begin{equation*}}
\newcommand{\ees}{\end{equation*}}
\newcommand{\beas}{\begin{eqnarray*}}
\newcommand{\eeas}{\end{eqnarray*}}

\makeatletter
\newtheorem*{rep@theorem}{\rep@title}
\newcommand{\newreptheorem}[2]{%
\newenvironment{rep#1}[1]{%
 \def\rep@title{#2 \ref{##1} (restated)}%
 \begin{rep@theorem}}%
 {\end{rep@theorem}}}
\makeatother

\allowdisplaybreaks

\newtheorem{thm}{Theorem}
\newtheorem*{thm*}{Theorem}
\newtheorem{cor}[thm]{Corollary}

\newtheorem{lem}[thm]{Lemma}
\newtheorem*{lem*}{Lemma}

\newtheorem{prop}[thm]{Proposition}

\newreptheorem{thm}{Theorem}
\newreptheorem{lem}{Lemma}
\newreptheorem{cor}{Corollary}

\newcommand{\cs}[1]{{\color{red}CS: #1}}

\begin{document}

\title{Quantum vs.\ classical algorithms for solving the heat equation}% Force line breaks with \\

\author{Noah Linden}
\email{n.linden@bristol.ac.uk}
\author{Ashley Montanaro}%
\email{ashley.montanaro@bristol.ac.uk}
\author{Changpeng Shao}
\email{changpeng.shao@bristol.ac.uk}
\affiliation{\vspace{-11pt}School of Mathematics, Fry Building, University of Bristol, UK}

\date{\today}

\begin{abstract}
%Partial differential equations play an important role in many areas of science and engineering.
Quantum computers are predicted to outperform classical ones for solving partial differential equations, perhaps exponentially. Here we consider a prototypical PDE -- the heat equation in a rectangular region -- and compare in detail the complexities of ten classical and quantum algorithms for solving it, in the sense of approximately computing the amount of heat in a given region. We find that, for spatial dimension $d \ge 2$, there is an at most quadratic quantum speedup using an approach based on applying amplitude estimation to an accelerated classical random walk. However, an alternative approach based on a quantum algorithm for linear equations is never faster than the best classical algorithms.
\end{abstract}

\maketitle

\setlength{\parskip}{3pt}

%-------------------------------------------------------------------------------

%\section{Introduction}

Quantum computers are predicted to solve certain problems substantially more efficiently than their classical counterparts. One area where quantum algorithms could significantly outperform classical ones is the approximate solution of partial differential equations (PDEs). This prospect is both exciting and plausible: exciting because of the ubiquity of PDEs in many fields of science and engineering, and plausible because some of the leading classical approaches to solving PDEs (e.g.\ via the finite difference or finite element methods) are based on discretising the PDE and reducing the problem to solving a system of linear equations. There are quantum algorithms that solve linear equations exponentially faster than classical algorithms (in a certain sense), via approaches that stem from the algorithm of Harrow, Hassidim and Lloyd (HHL)~\cite{harrow09}, so these algorithms could be applied to PDEs. There have been a succession of papers in this area which have developed new quantum algorithmic techniques~\cite{leyton08,clader13,berry14,berry17,arrazola19,childs20,lubasch19,childs20a,xin20} and applied quantum algorithms to particular problems~\cite{clader13,cao13,scherer17,wang19,costa19}.

However, in order to determine if a genuine quantum speedup can be obtained, it is essential to take into account all complexity parameters, and to compare against the best classical algorithms. The quantum algorithm should be given the same task as the classical algorithm -- to produce a classical solution to a classical problem, up to a certain level of accuracy -- rather than (for example) being asked to produce a quantum superposition corresponding to the solution. This can sometimes lead to apparently exponential speedups being reduced substantially. For example, it was suggested that quantum algorithms for the finite element method could solve electromagnetic scattering cross-section problems exponentially more efficiently than classical algorithms~\cite{clader13}, but it was later argued that the speedup can be at most polynomial~\cite{montanaro16b} (in fixed spatial dimension). The true extent of the achievable speedup (or otherwise) by quantum algorithms for PDEs over their classical counterparts remains to be seen.

Here we aim to fix a benchmark problem to enable us to compare the complexities of classical and quantum algorithms for solving PDEs. The analysis of~\cite{montanaro16b}, for example, was not specific to a particular problem, and also focused only on the finite element method; here, by contrast, we aim to choose a specific problem and pin down whether quantum algorithms of various forms can solve it more quickly than standard classical algorithms. We will consider the heat equation, which has a number of desirable features in this context: it is a canonical problem which has been studied extensively; it has many applications; and there are many methods known for solving it.

%-------------------------------------------------------------------------------

\subsection{Our results}

We compare the complexity of five classical methods and five quantum methods for solving the heat equation:
\be \label{eq:heatequation} 
\frac{\partial u}{\partial t} = \alpha \left( \frac{\partial^2 u}{\partial x_1^2} + \dots + \frac{\partial^2 u}{\partial x_d^2} \right) 
\ee
for some $\alpha > 0$, in $d$ spatial dimensions. We consider the hypercubic spatial region $x_i \in [0,L]$ and the time region $t \in [0,T]$, and let $R = [0,L]^d \times [0,T]$. We use periodic boundary conditions for each $x_i$, but not $t$. We fix the boundary conditions $u(x_1,\dots,x_d,0) = u_0(x_1,\dots,x_d)$ for some ``simple'' function $u_0:\R^d \rightarrow \R^{\ge 0}$ that is known in advance. We henceforth use boldface to denote vectors, and in particular let $\mathbf{x}$ denote the vector $(x_1,\dots,x_d)$. To get some intution for ``reasonable'' relationships between some of the parameters, $T \gg L^2/\alpha$ is a typical timescale for the distribution of heat to approach the uniform distribution.

In our bounds, we aim to compare the complexity of classical and quantum techniques for solving (\ref{eq:heatequation}), while avoiding a dependence on the complexity of $u_0$. Therefore, we assume that $u_0(x_1,\dots,x_d)$ can be computed exactly at no cost for all $x_1,\dots,x_d$, and further that $\int_{S} u_0(x_1,\dots,x_d) dx_1 \dots d{x_d}$ and $\int_{S} u_0^2(x_1,\dots,x_d) dx_1 \dots d{x_d}$ can be computed exactly at no cost for all regions $S$. Below, we will extend this assumption to being able to compute sums of simple functions of $u_0(x_1,\dots,x_d)$ values over discretised regions. (Note that all of the classical and quantum algorithms we consider have some requirement for an assumption of this form, so we are not giving one type of algorithm an unfair advantage over the other.)

%Assume that $\int_{[0,L]^d} u_0(x_1,\dots,x_d) dx_1 \dots d_{x_d} = 1$, which sets the overall normalisation for the problem, as the heat equation preserves the $L_1$ norm.
We will additionally assume that, for all $i,j \in \{1,\dots,d\}$ and some smoothness bound $\zeta$ of dimension (length)$^{-4}$ if $u$ is dimensionless,
\bea \max_{(x_1,\dots,x_d,t) \in R} \left|\frac{\partial^4 u}{\partial x_i^2 \partial x_j^2}(x_1,\dots,x_d,t) \right| \le \frac{\zeta}{L^d},\\
\max_{(x_1,\dots,x_d,t) \in R} \left|\frac{\partial^2 u}{\partial x_i^2}(x_1,\dots,x_d,t) \right| \le \frac{\zeta}{L^{d-2}},\\
\max_{(x_1,\dots,x_d,t) \in R} \left|\frac{\partial u}{\partial x_i}(x_1,\dots,x_d,t) \right| \le \frac{\zeta}{L^{d-3}}. \eea %\frac{\zeta}{L^4}
The denominators in these bounds are chosen to be appropriate based on dimensional analysis considerations; similar scaling for the second and first derivative bounds can be obtained directly from a bound on 4th derivatives and on $u$ itself~\cite{ore38}.

% (we will see later why this is a reasonable choice of constraint -- for now, note that division by $L^4$ makes $\zeta$ dimensionless).

There are many interpretations one could consider of what it means to ``solve'' the heat equation. Here we focus on solving the following problem: given $\epsilon \in (0,1)$, a fixed $t \in [0,T]$, and a subset $S \subseteq [0,L]^d$, output $\widetilde{H}$ such that
\be \label{eq:heatintegral} \left| \widetilde{H} - \int_S u(x_1,\dots,x_d,t) dx_1 \dots d{x_d} \right| \le \epsilon \ee
with probability at least 0.99. That is, for a given time, and a given spatial region, we aim to approximate the total amount of heat within that region. The complexity of solving the heat equation depends on the desired accuracy $\epsilon$ as well as all of the other parameters. We usually imagine that these other parameters are fixed first, and then consider the scaling of the complexity with respect to $\epsilon$.

All of the algorithms we studied were based on the standard approach of discretising the equation (\ref{eq:heatequation}) via the finite difference method, leading to a system of linear equations. Specifically, we used the simple ``forward time, central space'' (FTCS) method with a uniform rectangular grid. We evaluated the following classical algorithms:
\begin{itemize}
    \item Solving the corresponding system of linear equations using the conjugate gradient method.
    \item Iterating forward in time from the initial condition.
    \item Using the Fast Fourier Transform to solve the linear system.
    \item A random walk method based on the connection between the heat equation and random walk on a grid~\cite{lawler10,kac47,king51}.
    \item An accelerated version of the random walk method, using efficient sampling from the binomial distribution\footnote{A similar complexity can be achieved using a somewhat more complex approach based on the multilevel Monte Carlo method~\cite{cliffe11,giles08}.}.
\end{itemize}
We also evaluated the following quantum algorithms:
\begin{itemize}
    \item Solving the linear system using the fastest quantum algorithms for solving linear equations~\cite{chakraborty18}.
    \item Diagonalising the linear system using the quantum Fourier transform and postselection.
    \item Coherently accelerating the random walk on a grid~\cite{apers18,gilyen19}.
    \item Applying amplitude estimation~\cite{brassard02} to the classical random walk on a grid.
    \item Applying amplitude estimation to the fast classical random walk algorithm.
\end{itemize}
These methods vary in their flexibility. For example, the quantum and classical linear equations methods can be applied to much more general boundary conditions and spatial domains than those considered here (and to other PDEs), whereas the Fast Fourier Transform and coherent diagonalisation methods are only immediately applicable to solving the heat equation in a simple region.

There are still more solution methods that could be considered (e.g.\ the use of different discretisation techniques). One example is solving the heat equation by expressing it as a system of ODEs, by discretising only the right-hand side of (\ref{eq:heatequation}). A high-precision quantum algorithm for systems of ODEs was given in~\cite{berry17}. However, applying it to the heat equation seems to give a complexity somewhat worse than solving the fully discretised system of linear equations using a quantum algorithm (see Appendix \ref{app:odealgm}). One can also solve the heat equation in the specific case of a hyperrectangular region by using the known explicit solution in terms of Fourier series. This requires computing integrals dependent on the initial condition $u_0$, but for certain initial conditions, it may be more efficient (or even give an exact solution).

Our results are summarised in Table \ref{fig:table}, where we display runtimes in terms of $\epsilon$ alone, although we compute the complexity of the various algorithms in terms of the other parameters in detail below. The key points are as follows:

\begin{itemize}
    \item For $d =1$, the quantum methods are all outperformed by the classical Fast Fourier Transform method. For $d \ge 2$, the fastest method is the quantum algorithm based on applying amplitude amplification to a ``fast'' classical random walk. For arbitrary $d$, the largest quantum speedup using this method is from $\widetilde{O}(\epsilon^{-2})$ to $\widetilde{O}(\epsilon^{-1})$.
    \item The Fast Fourier Transform and fast random walk amplitude estimation algorithms are specific to a rectangular region. Considering algorithms that could also be applied to more general regions, the fastest classical method for $d \le 3$ is iterating the initial condition forward in time. This outperforms all quantum methods in $d=1$, performs roughly as well as (standard) random walk amplitude estimation in $d=2$, and is outperformed by random walk amplitude estimation for $d \ge 3$.
    \item The quantum linear equation solving method is always outperformed by other quantum methods. However, note that it provides more flexibility in terms of estimating other quantities, and allowing for different boundary conditions.
    \item Among the {\em space-efficient} methods -- those which use space polylogarithmic in $1/\epsilon$ -- there is a quantum speedup in all dimensions (from $\widetilde{O}(\epsilon^{-2})$ to $\widetilde{O}(\epsilon^{-1})$), because this criterion rules out the classical Fast Fourier Transform method.
\end{itemize}

These bounds do not assume the use of a preconditioner to improve the condition number of the relevant linear system. If a perfect preconditioner were available, then the complexity of the quantum linear equation solving method would be reduced to be comparable with that of the diagonalisation method, but would still not be competitive with other methods.

We conclude that, if our results for the heat equation are representative of the situation for more general PDEs, it is unclear whether quantum algorithms will offer a super-polynomial advantage over their classical counterparts for solving PDEs, but polynomial speedups may be available.

\begin{table*}[t]
\[
\begin{array}{|c|c|c|c|c|c|c|c|}
\hline
& \text{Method} & \text{Region} & \text{Thm.} & d=1 & d=2 & d=3 & d\ge 4\\
\hline
\text{Classical} & \text{* Linear equations} & \text{General} & \ref{thm:lineqmethod} & \widetilde{O}(\epsilon^{-2}) & \widetilde{O}(\epsilon^{-2.5}) & \widetilde{O}(\epsilon^{-3}) & \widetilde{O}(\epsilon^{-d/2-1.5}) \\
& \text{* Time-stepping} & \text{General} & \ref{thm:simplelineqmethod} & \widetilde{O}(\epsilon^{-1.5}) & \widetilde{O}(\epsilon^{-2}) & \widetilde{O}(\epsilon^{-2.5}) & \widetilde{O}(\epsilon^{-d/2-1})\\
& \text{* Fast Fourier Transform} & \text{Rectangular} & \ref{thm:fftmethod} & \mathbf{\widetilde{O}(\epsilon^{-0.5})} & \widetilde{O}(\epsilon^{-1}) & \widetilde{O}(\epsilon^{-1.5}) & \widetilde{O}(\epsilon^{-d/2}) \\
& \text{Random walk} & \text{General} & \ref{thm:rwmethodest} & \widetilde{O}(\epsilon^{-3}) & \widetilde{O}(\epsilon^{-3}) & \widetilde{O}(\epsilon^{-3}) & \widetilde{O}(\epsilon^{-3}) \\
& \text{Fast random walk} & \text{Rectangular} & \ref{thm:fastrwmethodest} & \widetilde{O}(\epsilon^{-2}) & \widetilde{O}(\epsilon^{-2}) & \widetilde{O}(\epsilon^{-2}) & \widetilde{O}(\epsilon^{-2}) \\
\hline
\text{Quantum} & \text{Linear equations} & \text{General} & \ref{lem:qprobest} & \widetilde{O}(\epsilon^{-2.5}) & \widetilde{O}(\epsilon^{-2.5})& \widetilde{O}(\epsilon^{-2.75}) & \widetilde{O}(\epsilon^{-d/4-2}) \\
 & \text{Coherent random walk acceleration} & \text{General} & \ref{thm:ffrwmethod} & \widetilde{O}(\epsilon^{-1.75}) & \widetilde{O}(\epsilon^{-2}) & \widetilde{O}(\epsilon^{-2.25}) & \widetilde{O}(\epsilon^{-d/4-1.5}) \\
& \text{Coherent diagonalisation} & \text{Rectangular} & \ref{thm:diag} & \widetilde{O}(\epsilon^{-1.25}) & \widetilde{O}(\epsilon^{-1.5}) & \widetilde{O}(\epsilon^{-1.75}) & \widetilde{O}(\epsilon^{-d/4-1}) \\
& \text{Random walk amplitude estimation} & \text{General} & \ref{thm:accelrwmethodest} & \widetilde{O}(\epsilon^{-2}) & \widetilde{O}(\epsilon^{-2}) & \widetilde{O}(\epsilon^{-2}) & \widetilde{O}(\epsilon^{-2})\\
& \text{Fast r.w.\ amplitude estimation} & \text{Rectangular} & \ref{thm:fastaccelrwmethodest} & \widetilde{O}(\epsilon^{-1}) & \mathbf{\widetilde{O}(\epsilon^{-1})} & \mathbf{\widetilde{O}(\epsilon^{-1})} & \mathbf{\widetilde{O}(\epsilon^{-1})} \\
\hline
\end{array}
\]
\caption{The runtimes of the various algorithms considered in this work for solving the heat equation up to accuracy $\epsilon$ in spatial dimension $d$, in terms of $\epsilon$ and $d$ only. $\widetilde{O}$ notation hides polylogarithmic factors. Lowest-complexity algorithms for each $d$ highlighted in bold. Starred methods use $\poly(1/\epsilon)$ space; other methods use $\poly\log(1/\epsilon)$ space.}
\label{fig:table}
\end{table*}

In the remainder of this work, we prove the results corresponding to the complexities reported in Table \ref{fig:table}. We begin by describing the discretisation and numerical integration approach used, before going on to describe and determine the complexity of the various algorithms. To achieve this, we need to obtain several technical bounds (e.g.\ on the condition number of the relevant linear system; on the $\ell_2$ norm of a solution to the heat equation; and on the complexity of approximating the heat in a region from a quantum state corresponding to a solution to the heat equation). We aim for a self-contained presentation wherever possible, rather than referring to results in the extensive literature on numerical solutions of PDEs; see \cite{iserles09,leveque07,trefethen96} for further details.

%-------------------------------------------------------------------------------

\section{Technical ingredients}

In this section we will discuss the key ingredients that are required for quantum and classical algorithms to solve the heat equation.

%-------------------------------------------------------------------------------

\subsection{Discretisation}
\label{sec:discretisation}

All of the algorithms that we will consider are based on discretising the PDE (\ref{eq:heatequation}). Here we will consider the simplest method of discretisation, known as the forward-time, central-space (FTCS) method. This method is based on discretising using the following equalities (for one variable), which can be proved using Taylor's theorem with remainder:
\be \label{eq:dxapprox} \frac{d u}{d t} = \frac{u(t+h) - u(t)}{h} - \frac{h}{2} \frac{d^2u}{dt^2}(\xi) \ee
%\be \label{eq:dx2approx} \frac{\partial^2 u}{\partial x^2} = \frac{u(x+h) + u(x-h) - 2u(x)}{h^2} \pm C h^2 \ee
\be \label{eq:dx2approx} \frac{d^2 u}{d x^2} = \frac{u(x+h) + u(x-h) - 2u(x)}{h^2} + \frac{h^2}{24}\left(\frac{d^4 u}{dx^4}(\xi') + \frac{d^4 u}{dx^4}(\xi'') \right), \ee
where we assume that $u$ is 4 times differentiable, and $\xi \in [t,t+h]$, $\xi' \in [x,x+h]$, $\xi'' \in [x-h,x]$. So
\be \label{eq:approx1} \left| \frac{d u}{d t} - \frac{u(t+h) - u(t)}{h} \right| \le  \frac{h}{2} \sup_t \left|\frac{d^2 u}{dt^2}(t)\right| \ee
\be \label{eq:approx2} \left| \frac{d^2 u}{d x^2} - \frac{u(x+h) + u(x-h) - 2u(x)}{h^2} \right| \le \frac{h^2}{12} \sup_x \left|\frac{d^4 u}{dx^4}(x)\right|. \ee
We will apply these approximations to multivariate functions $u(\mathbf{x},t)$ that satisfy, for all $i,j \in \{1,\dots,d\}$, 
\be \max_{(x_1,\dots,x_d,t) \in R} \left|\frac{\partial^4 u}{\partial x_i^2 \partial x_j^2}(x_1,\dots,x_d,t) \right| \le \frac{\zeta}{L^d} \ee %\frac{\zeta}{L^4} \ee
for some $\zeta$ and all $(\mathbf{x},t) \in R$. From (\ref{eq:heatequation}), this implies that $\max_{(x_1,\dots,x_d,t) \in R}|\frac{\partial^2 u}{\partial t^2}(\mathbf{x},t)| \le \zeta \alpha^2 d^2 / L^d$. We note that this is dimensionally consistent as $\alpha$ has dimensions (length)$^2/$time and $u$ is a density.

We will use the sequence of discrete positions $x_0 = 0, x_1 = \Delta x, \dots, x_n = n \Delta x$; $t_0 = 0, t_1 = \Delta t, \dots, t_m = m \Delta t$, such that $T = m \Delta t$, $L = n \Delta x$. Let $G$ (for ``grid'') denote the set of points $(\mathbf{x},t) \in R$ such that the coordinates of $\mathbf{x}$ are integer multiples of $\Delta x$, and $t$ is an integer multiple of $\Delta t$. We will let the vector $\mathbf{u}$ denote the exact solution of (\ref{eq:heatequation}) at points in $G$, and will use $\widetilde{u}$ or $\mathbf{\widetilde{u}}$ for the approximate solution to (\ref{eq:heatequation}) found via discretisation, dependent on whether we are considering this as a function or a vector.

Considering points in $G$ and using the approximations (\ref{eq:approx1}) and (\ref{eq:approx2}) gives the linear constraints
\be \label{eq:lincons} \frac{\widetilde{u}(\mathbf{x},t+\Delta t) - \widetilde{u}(\mathbf{x},t)}{\Delta t} = \frac{\alpha}{\Delta x^2} \sum_{i=1}^d 
\Big(
\widetilde{u}(\dots,x_i+\Delta x,\dots,t) + \widetilde{u}(\dots,x_i-\Delta x,\dots,t) - 2 \widetilde{u}(\mathbf{x},t)\Big).
\ee
The following result can be shown using standard techniques.

\begin{thm}[Approximation up to small $\ell_\infty$ error]
\label{thm:approxlinfty}
If $\Delta t \le \Delta x^2/(2d\alpha)$,
\be \|\mathbf{\widetilde{u}} - \mathbf{u}\|_\infty \le \frac{\zeta \alpha d T}{L^d} \left(\frac{\alpha d \Delta t}{2} + \frac{\Delta x^2}{12} \right). \ee
\end{thm}

\begin{proof}
From (\ref{eq:lincons}),
\be
\label{eq:linstep} \widetilde{u}(\mathbf{x},t+\Delta t) = \left(1 - \frac{2d\alpha \Delta t}{\Delta x^2} \right) \widetilde{u}(\mathbf{x},t) + \frac{\alpha \Delta t}{\Delta x^2} \sum_{i=1}^d \Big(\widetilde{u}(\dots,x_i+\Delta x,\dots,t) + \widetilde{u}(\dots,x_i-\Delta x,\dots,t)\Big).
\ee
Let $\mathcal{L}$ be the linear operator defined by the right-hand side of (\ref{eq:linstep}). Letting $\mathbf{\widetilde{u}_i}$ and $\mathbf{u_i}$ denote the approximate and exact solutions at time $t_i$ (i.e.\ the $n^d$-component vectors $\widetilde{u}(\cdot,t_i)$, $u(\cdot,t_i)$), we have $\mathbf{\widetilde{u}_{i+1}} = \mathcal{L} \mathbf{\widetilde{u}_i}$. $\mathcal{L}$ is stochastic if
\be \label{eq:dtdx} 1 - \frac{2d \alpha \Delta t}{\Delta x^2} \ge 0,\;\;\;\; \text{i.e.}\;\;\;\; \Delta t \le \frac{\Delta x^2}{2d\alpha}, \ee
and this condition holds by assumption. By the discretisation error bounds (\ref{eq:approx1}), (\ref{eq:approx2}),
\bea 
\nonumber && \left| \frac{u(\mathbf{x},t+\Delta t) - u(\mathbf{x},t)}{\Delta t} - \frac{\alpha}{\Delta x^2}  \sum_{i=1}^d 
\Big( u(\dots,x_i+\Delta x,\dots,t) + u(\dots,x_i-\Delta x,\dots,t) - 2u(\mathbf{x},t)\Big) \right| \\
&\le& \frac{\zeta}{L^d}\left(\frac{\alpha^2 d^2 \Delta t}{2} + \frac{\alpha d \Delta x^2}{12}\right), 
\eea
implying
\bea
\nonumber && \hspace{-.5cm} \left| u(\mathbf{x},t+\Delta t) - \left( \left(1 - \frac{2d\alpha \Delta t}{\Delta x^2} \right) u(\mathbf{x},t) +  \frac{\alpha \Delta t}{\Delta x^2} \sum_{i=1}^d u(\dots,x_i+\Delta x,\dots,t) + u(\dots,x_i-\Delta x,\dots,t) \right) \right| \\
&\le& \frac{\zeta \alpha d \Delta t}{L^d} \left(\frac{\alpha d \Delta t}{2} + \frac{\Delta x^2}{12}\right),
\eea
i.e.
\be\label{bound of infty norm}
\| \mathbf{u_{i+1}} - \mathcal{L} \mathbf{u_i} \|_\infty \le \frac{\zeta \alpha d \Delta t}{L^d} \left(\frac{\alpha d \Delta t}{2} + \frac{\Delta x^2}{12}\right). 
\ee
Writing $\mathbf{\widetilde{u}_i} = \mathbf{u_i} + \mathbf{e_i}$ for some error vector $\mathbf{e_i}$, we have
\bea \mathbf{\widetilde{u}_0} &=& \mathbf{u_0}\\
 \mathbf{\widetilde{u}_1} &=& \mathcal{L} \mathbf{u_0} = \mathbf{u_1} + \mathbf{e_1},\;\;\;\; \text{where } \|\mathbf{e_1}\|_\infty \le \frac{\zeta \alpha d \Delta t}{L^d} \left(\frac{\alpha d \Delta t}{2} + \frac{\Delta x^2}{12}\right) \\
 \mathbf{\widetilde{u}_2} &=& \mathcal{L}\mathbf{\widetilde{u}_1} = \mathcal{L}(\mathbf{u_1} + \mathbf{e_1}) = \mathbf{u_2} + \mathbf{e_2} + \mathcal{ L}\mathbf{e_1},\;\;\;\; \text{where } \|\mathbf{e_2}\|_\infty \le \frac{\zeta \alpha d \Delta t}{L^d} \left(\frac{\alpha d \Delta t}{2} + \frac{\Delta x^2}{12}\right); \eea
as $\mathcal{L}$ is stochastic, $\|\mathcal{L}\mathbf{e_1}\|_\infty \le \|\mathbf{e_1}\|_\infty$, so $\|\mathbf{\widetilde{u}_2} - \mathbf{u_2}\|_\infty \le 2\zeta \alpha d \Delta t L^{-d} \left(\frac{\alpha d \Delta t}{2} + \frac{\Delta x^2}{12}\right)$. Repeating this argument,
\be \| \mathbf{\widetilde{u}_m} - \mathbf{u_m}\|_\infty \le \frac{m \zeta \alpha d \Delta t}{L^d} \left(\frac{\alpha d \Delta t}{2} + \frac{\Delta x^2}{12}\right) = \frac{\zeta \alpha d T}{L^d} \left(\frac{\alpha d \Delta t}{2} + \frac{\Delta x^2}{12} \right) \ee
as claimed.
\end{proof}

\begin{cor}
\label{cor:accuracy}
To estimate $\mathbf{u}$ up to $\ell_\infty$ accuracy $\epsilon/L^d$, it is sufficient to take
\be \Delta t = \frac{3\epsilon}{2d^2\alpha^2\zeta T},\;\;\;\; \Delta x = \sqrt{\frac{3\epsilon}{d\alpha\zeta T}}. \ee
This corresponds to taking $m = 2T^2 d^2 \alpha^2  \zeta / (3\epsilon)$, $n = L \sqrt{d \alpha \zeta T/ (3\epsilon)}$.
\end{cor}

\begin{proof}
By design, $\Delta t = \Delta x^2 / (2d\alpha)$, so Theorem \ref{thm:approxlinfty} can be applied. Insertion of the stated values into Theorem \ref{thm:approxlinfty} gives the claimed result.
\end{proof}

Note that the constant factors in $\Delta t$ and $\Delta x$ could be traded off against one another to some extent, and that the constraint that spatial 4th derivatives are upper-bounded by $\zeta/L^d$ applies to the solution $u$ to the heat equation, rather than the initial condition $u_0$. However, for any $t$, $\|\frac{\partial^4u}{\partial x_i^4}(\mathbf{x},t)\|_\infty \le \|\frac{\partial^4u_0}{\partial x_i^4}(\mathbf{x})\|_\infty$, so such a constraint on $u_0$ implies an equivalent constraint on $u$ at other times $t$. (This claim follows from the discretisation argument of Theorem \ref{thm:approxlinfty}: the linear time-evolution operator $\mathcal{L}$ defined in the theorem cannot increase the infinity-norm, and discretised partial-derivative operators commute with $\mathcal{L}$.)

We will make the choices for $m$ and $n$ specified in Corollary \ref{cor:accuracy} throughout the rest of the paper. Observe that, with these choices, the operator $\mathcal{L}$ is precisely a simple random walk on $\Z_n^d$.

Now we have introduced the discretisation method, we can describe the normalisation used: we assume that
\be \|\mathbf{u_0}\|_1 = \sum_{(\mathbf{x},0) \in G} u_0(\mathbf{x}) = \left(\frac{n}{L}\right)^d = \Delta x^{-d}. \ee
By stochasticity of $\mathcal{L}$, this implies that $\|\mathbf{\widetilde{u}_i}\|_1 = \Delta x^{-d}$ for all $i$. This assumption is approximately equivalent to assuming that $\int_{[0,L]^d} u_0(\mathbf{x}) dx_1 \dots dx_d = 1$; we will discuss why at the end of the next section. As a quick check, note that taking $u_0(\mathbf{x}) = L^{-d}$ gives $\|\mathbf{u_0}\|_1 = \left(\frac{n}{L}\right)^d$, $\int_{[0,L]^d} u_0(\mathbf{x}) dx_1 \dots dx_d = 1$.

%-------------------------------------------------------------------------------

\subsection{Numerical integration}
\label{sec:numint}

Our goal will ultimately be to compute the integral defined in (\ref{eq:heatintegral}) giving the total amount of heat within a region $S$ approximately, at a fixed time. Following the discretisation approach, we will have access to (approximate) evaluations of a function $u$ at equally spaced grid points, and seek to compute the integral of $u$ over $S$.

We will consider several numerical integration methods for achieving this goal. Each of them is based on a 1-dimensional approximation of the form
\be \int_a^b f(x) dx = \Delta x \sum_i w(i) f(x_i) + E, \ee
where $w(i)$ are real weights, $x_i$ are grid points between $a$ and $b$ with spacing $\Delta x$, where $b-a$ is an integer multiple of $\Delta x$, and $E$ is an error term. If we define $\mathbf{w}$, $\mathbf{f}$ to be the vectors corresponding to evaluations of $w$ and $f$ at grid points, we can write the approximation as $\Delta x \mathbf{w} \cdot \mathbf{f}$. To extend an approximation of this form to $d$-variate functions, we simply apply it in each dimension, e.g.\ for $d=2$:
\bea
\int_{a_1}^{b_1} \int_{a_2}^{b_2} f(x,y) dy dx &=& \int_{a_1}^{b_1} \left( \Delta x \sum_i w(i) f(x,y_i) + E(x) \right) dx\\
&=& \Delta x \sum_i w(i) \int_{a_1}^{b_1} f(x,y_i) dx + E' \\
&=& \Delta x \left( \sum_i w(i) \left( \Delta x \sum_j w(j) f(x_j,y_i) + E(i) \right) \right) + E'\\
&=& (\Delta x)^2 \sum_{i,j} w(i) w(j) f(x_j,y_i) + \Delta x \left( \sum_i w(i) E(i)\right) + E',
%&=& \frac{\Delta x \Delta y}{9} \big( (f(x_0,y_0) + 4f(x_0,y_1) + \dots + f(x_0,y_n))\\
%&+& 4(f(x_1,y_0) + 4f(x_1,y_1) + \dots + f(x_1,y_n))\\
%&+& \dots + (f(x_n,y_0) + 4f(x_n,y_1) + \dots + f(x_n,y_n))\big) + \frac{\Delta y}{3} E_x + E'_y,
\eea
where $E(x)$ is the error term for $x$, and $|E'| \le (b_1-a_1) \max_x |E(x)| \le L \max_x |E(x)|$.
For arbitrary $d$, it is straightforward to see that we can interpret this approximation as computing the inner product $(\Delta x)^d \mathbf{w}^{\otimes d} \cdot \mathbf{f}$. The error bound becomes $O(d L^{d-1} \max_x |E(x)|)$, as we will always have $\sum_i w(i) \le n$.

When applied to the heat equation, we seek to evaluate $\int_S u(\mathbf{x},t) d\mathbf{x}$ for some subset $S \subseteq [0,L]^d$ and a fixed time $t$.
Applying the above approximation gives a weighted sum of the form
\be \label{eq:genapproxint} (\Delta x)^d \sum_{(\mathbf{x},t) \in G \cap S} w(\mathbf{x}) \widetilde{u}(\mathbf{x},t), \ee
where $G$ is a set of grid points of spacing $\Delta x$.
Then
\beas
&& \left| \sum_{(\mathbf{x},t) \in G \cap S} (\Delta x)^d w(\mathbf{x}) \widetilde{u}(\mathbf{x},t) - \int_S u(\mathbf{x},t) d\mathbf{x} \right|\\
 &\le& (\Delta x)^d \left| \sum_{(\mathbf{x},t) \in G \cap S } w(\mathbf{x}) \widetilde{u}(\mathbf{x},t) - \sum_{(\mathbf{x},t) \in G \cap S} w(\mathbf{x}) u(\mathbf{x},t) \right| + \left| \sum_{(\mathbf{x},t) \in G \cap S} (\Delta x)^d w(\mathbf{x}) u(\mathbf{x},t) - \int_S u(\mathbf{x},t) d\mathbf{x} \right|\\
 &\le& (\Delta x)^d \sum_{(\mathbf{x},t) \in G \cap S} w(\mathbf{x}) \left| \widetilde{u}(\mathbf{x},t) - u(\mathbf{x},t) \right| + d L^{d-1} E \\
  &\le& (\Delta x)^d \|\mathbf{w}\|_1^d \zeta \alpha d T L^{-d} \left(\frac{\alpha d \Delta t}{2} + \frac{\Delta x^2}{12} \right) + d L^{d-1} E,% \le C'' \epsilon.
\eeas
where $E = \max_x E(x)$, the second inequality follows from the previous error analysis, and the final inequality follows from Theorem \ref{thm:approxlinfty}. As $\Delta t = \Delta x^2 / (2d\alpha)$ from Corollary \ref{cor:accuracy}, this corresponds to a bound which is
\be \label{eq:generalbound} O((\Delta x)^{d+2}\|\mathbf{w}\|_1^d L^{-d}\alpha d \zeta T + d L^{d-1} E). \ee
We will consider three numerical integration methods that fit into the above framework:

\begin{enumerate}
    \item Simpson's rule: $x_i = a+i\Delta x$, $a \le x_i \le b$, $\mathbf{w} = \frac{1}{3}(1,4,2,4,2,\dots,4,1)$,
    \be |E| \le \frac{\Delta x^4}{180} (b-a) \max_{\xi \in [a,b]} \left|\frac{d^4f}{dx^4}(\xi) \right|. \ee
    Inserting into (\ref{eq:generalbound}) and using $|b-a| \le L$, $\|\mathbf{w}\|_1 \le n = L/\Delta x$, we obtain an overall error bound of
    \be O(\Delta x^2 \alpha d \zeta T+ d \Delta x^4 \zeta) = O(d \Delta x^2 \zeta(\alpha T + \Delta x^2)). \ee
    Assuming that $\Delta x \rightarrow 0$, the second term is negligible. Choosing $\Delta x$ as in Corollary \ref{cor:accuracy}, the final error introduced by numerical integration is $O(\epsilon)$.
    
    \item The midpoint rule: $x_i = a+(i+\frac12)\Delta x$, $a < x_i < b$, $\mathbf{w} = (1,1,\dots,1)$,
    \be |E| \le \frac{\Delta x^2}{24} (b-a) \max_{\xi \in [a,b]} \left|\frac{d^2f}{dx^2}(\xi) \right| = O(\Delta x^2 L^{3-d} \zeta). \ee
    Using a similar argument to the previous point, we obtain an overall error bound of
    \be O(\Delta x^2 \alpha d \zeta T + d \Delta x^2 L^2 \zeta) = O(d \Delta x^2 \zeta(\alpha T+ L^2)). \ee
    The error increases with $L$, so we may need to choose $\Delta x$ smaller than the choice made in Corollary \ref{cor:accuracy}. Indeed, working through the same argument, we obtain
    \be \label{eq:midpointmn} m = O(T\alpha d^2 \zeta(\alpha T + L^2)/\epsilon),\;\;\;\; n = O(L \sqrt{d\zeta(\alpha T + L^2)/\epsilon}). \ee
    However, for fixed $\alpha,d,T,L$ the asymptotic scaling is the same as Simpson's rule, and we will see below that this technique can be advantageous in two respects: the $\ell_2$ and $\ell_\infty$ norms of $\mathbf{w}$ are lower, and its values are all equal.
    
    \item The left Riemann sum: $x_i = a+i\Delta x$, $a \le x_i < b$, $\mathbf{w} = (1,1,\dots,1)$,
    \be |E| \le \frac{\Delta x}{2} (b-a) \max_{\xi \in [a,b]} \left|\frac{df}{dx}(\xi) \right| = O(\Delta x L^{4-d} \zeta). \ee
    By the same argument, we obtain an overall error bound of
    \be O(\Delta x^2 \alpha d \zeta T + d \Delta x L^3 \zeta) = O(d \Delta x \zeta(\Delta x \alpha T + L^3)). \ee
    This is weaker than both of the previous bounds, but allows us to justify the normalisation assumption that we made that $\sum_{(\mathbf{x},0) \cap G} u_0(\mathbf{x}) = (\Delta x)^{-d}$. This is equivalent to the approximate integral of $u_0$ using the left Riemann sum in (\ref{eq:genapproxint}) equalling 1, which implies that for $\Delta x \rightarrow 0$, $\int_{\mathbf{x} \in [0,L]^d} u_0(\mathbf{x}) d\mathbf{x} \rightarrow 1$.
    
\end{enumerate}

%-------------------------------------------------------------------------------

\subsection{Condition number}

Since $\mathbf{\widetilde{u}_{i+1}} = \mathcal{L} \mathbf{\widetilde{u}_i}$ holds for $i=0,1,\ldots,m-1$, we can find a full approximate solution to the heat equation at all points in $G$ by solving the following linear system:
\be \label{linear system:forward method}
\left(
  \begin{array}{cccccc}
    I &   &        &&     \\
    -\mathcal{L}  & I           \\
       &     \ddots  & \ddots  &     \\
      &        &  -\mathcal{L} & I
  \end{array}
\right)
\left(
  \begin{array}{c}
    \mathbf{\widetilde{u}_1}     \\
    \mathbf{\widetilde{u}_2}           \\
    \vdots     \\
    \mathbf{\widetilde{u}_m}
  \end{array}
\right)
=\left(
  \begin{array}{c}
    \mathcal{L}\mathbf{\widetilde{u}_0}     \\
    0           \\
    \vdots     \\
    0
  \end{array}
\right).
\ee
An important quantity that determines the complexity of classical and quantum algorithms for solving a linear system $A \mathbf{x} = \mathbf{b}$ is the condition number $\kappa = \|A\|\|A^{-1}\|$. The proof of the following theorem is given in Appendix \ref{appendix:Estimation of the condition number}.

\begin{thm}
\label{thm:condition}
The matrix $A$ in (\ref{linear system:forward method}) satisfies $\|A\| = \Theta(1)$, $\|A^{-1}\| = \Theta(m)$. Hence the condition number is $\Theta(m)$.
\end{thm}

Also note that $\mathcal{L}$ appears on the right-hand side of (\ref{linear system:forward method}), raising the question of the complexity of preparing the vector (or quantum state) $\mathcal{L}\mathbf{\widetilde{u}_0} = \mathcal{L}\mathbf{u_0}$. In the quantum case, this complexity depends on the condition number of $\mathcal{L}$, which in general could be high; indeed, $\mathcal{L}$ can sometimes be noninvertible. However, we have made the assumption that the initial vector $\mathbf{u_0}$ is non-negative, and for all vectors of this form, $\mathcal{L}$ is well-conditioned:

\begin{lem}
\label{lem:conditiononpositive}
Let $\mathcal{L}$ be defined by (\ref{eq:linstep}), taking $\Delta t = \Delta x^2/(2\alpha d)$ as in Corollary \ref{cor:accuracy}. Then for all nonnegative vectors $\mathbf{u}$, $\|\mathcal{L}\mathbf{u}\|_2^2 / \|\mathbf{u}\|_2^2 \ge 1/(2d)$.
\end{lem}

The proof is included in Appendix \ref{app:conditiononpositive}.

%-------------------------------------------------------------------------------

\section{Classical methods}

Next we determine the complexity of various classical methods for solving the heat equation, based on the analysis of the previous section.

%-------------------------------------------------------------------------------

\subsection{Linear systems}
\label{sec:linear}

A standard classical method for the heat equation (and more general PDEs) is simply to solve the system of linear equations defined in Section \ref{sec:discretisation} directly. A leading approach for solving sparse systems of linear equations is the conjugate gradient method~\cite{shewchuk94}. This can solve a system of $N$ linear equations, each containing at most $s$ unknowns, and corresponding to a matrix $A$ with condition number $\kappa$, up to accuracy $\delta$ in the energy norm $\|\cdot\|_A$ in time $O(s \sqrt{\kappa} N \log(1/\delta))$. The energy norm $\|\mathbf{x}\|_A$ with respect to a positive semidefinite matrix $A$ is defined as $\|\mathbf{x}\|_A = \sqrt{\mathbf{x}^T A \mathbf{x}}$.

Note that as the dependence on $1/\delta$ is logarithmic, using almost any reasonable norm would not change this complexity bound much. For example, we have
\be \label{eq:2normvsanorm} \| \widetilde{\mathbf{x}} - \mathbf{x} \|_2 = \| A^{-1/2} A^{1/2} (\widetilde{\mathbf{x}} - \mathbf{x}) \|_2 \le \|A^{-1/2}\| \| A^{1/2} (\widetilde{\mathbf{x}} - \mathbf{x}) \|_2 = \|A^{-1}\|^{1/2} \| \widetilde{\mathbf{x}} - \mathbf{x} \|_A, \ee
where $\|\cdot\|$ denotes the operator norm.

\begin{thm}[Classical linear equations method]
\label{thm:lineqmethod}
There is a classical algorithm that outputs an approximate solution $\widetilde{u}(\mathbf{x},t)$ such that $|\widetilde{u}(\mathbf{x},t) - u(\mathbf{x},t)| \le \epsilon/L^d$ for all $(\mathbf{x},t) \in G$ in time
\be O\left(3^{-d/2} T^{d/2+3} L^d \left(\frac{\zeta}{\epsilon}\right)^{d/2+3/2} d^{d/2+4} \alpha^{d/2+3} \log(Td\alpha\zeta^{1/2}/\epsilon) \right). \ee
%$O(T^{4+\frac{d}{2}} L^d (d\alpha)^{4+\frac{d}{2}} (\zeta/\epsilon)^{2+\frac{d}{2}} \log(1/\epsilon))$.
\end{thm}

\begin{proof}
By Corollary \ref{cor:accuracy} and Theorem \ref{thm:condition}, we can achieve discretisation accuracy $\epsilon/L^d$ in the $\infty$-norm (which is sufficient to compute the amount of heat within a region up to accuracy $\epsilon$ via numerical integration) with a system of $N=O(mn^d)$ linear equations, each containing $O(d)$ variables, with condition number $\Theta(m)$, where $m = 2T^2 d^2 \alpha^2  \zeta / (3\epsilon)$, $n = L \sqrt{d \alpha \zeta T/ (3\epsilon)}$. We can also calculate the vector on the right-hand side of (\ref{linear system:forward method}) in time $O(dn^d)$ by multiplying $\mathbf{u_0}$ by $\mathcal{L}$.
Using the conjugate gradient method, this system can be solved up to accuracy $\delta$ in the energy norm in time $O(d m^{3/2} n^d \log(1/\delta))$. Then, by (\ref{eq:2normvsanorm}) and Theorem \ref{thm:condition}, to achieve accuracy $\epsilon$ in the $\ell_2$ norm (and hence the $\ell_\infty$ norm) it is sufficient to take $\delta = \Theta(\epsilon/\sqrt{m})$, giving an overall complexity of $O(d m^{3/2} n^d \log(m/\epsilon))$. Inserting the expressions for $m$ and $n$ gives the claimed result.
\end{proof}

The above approach based on linear equations can be used both for the forwards-in-time and backwards-in-time discretisation methods, and indeed to solve much more general PDEs than the heat equation. In the case of the forwards-in-time approach which is our focus here, there is an even simpler method: compute $\mathcal{L}^m \mathbf{u_0}$.

\begin{thm}[Classical time-stepping method]
\label{thm:simplelineqmethod}
There is a classical algorithm that outputs an approximate solution $\widetilde{u}(\mathbf{x},t)$ such that $|\widetilde{u}(\mathbf{x},t) - u(\mathbf{x},t)| \le \epsilon/L^d$ for all $(\mathbf{x},t) \in G$ in time $O(3^{-d/2} T^{{d}/{2}+2}L^d \alpha^{d/2+2} d^{d/2+3} (\zeta/\epsilon)^{d/2+1})$.% If $T,L,\alpha,d,\zeta=O(1)$, the algorithm runs in time $O(\epsilon^{-(d/2+1)})$.
\end{thm}

\begin{proof}
We simply apply the linear operator $\mathcal{L}$ defined in (\ref{eq:linstep}) $m$ times to the initial vector $\mathbf{u_0}$. Each matrix-vector multiplication can be carried out in time $O(dn^d)$, so all required vectors $\mathbf{\widetilde{u}_i}$ can be produced in $O(dmn^d)$ steps. Inserting the bounds for $m$ and $n$ from Corollary \ref{cor:accuracy} gives the claimed result.
\end{proof}

The time-evolution method described in Theorem \ref{thm:simplelineqmethod} is simple and efficient; however, the method of Theorem \ref{thm:lineqmethod} based on solving a full system of linear equations is more flexible. A natural alternative approach to compute $\mathcal{L}^\tau \mathbf{u_0}$ for some integer $\tau$ is to use the fast Fourier transform to diagonalise $\mathcal{L}$.

We will first need a technical lemma, which will also be used later on, about the complexity of computing eigenvalues of $\mathcal{L}^\tau$.

\begin{lem}
\label{lem:computeevs}
For any $\tau \in \{0,\dots,m\}$, and any $\delta > 0$, each eigenvalue of $\mathcal{L}^\tau$ can be computed up to accuracy $\delta$ in time $O(\log^2 (\tau/\delta)(\log \log (\tau/\delta) + \log \tau))$.
\end{lem}

\begin{proof}
It is shown in (\ref{eigenvalues of H}) and (\ref{eq:generall}) that
\be
\label{eq:lform}
\mathcal{L} =  I_{n}^{\otimes d} +
\frac{\alpha \Delta t}{\Delta x^2} 
\sum_{j=1}^d 
I_{n}^{\otimes (j-1)} \otimes H \otimes I_{n}^{\otimes (d-j)},
\ee
where $H$ is a circulant matrix with eigenvalues
\be \label{eq:lambdaj} \lambda_j = -4 \sin^2 \frac{j\pi}{n} \ee
for $j \in \{0,\dots,n-1\}$. Eigenvalues of $\mathcal{L}$ can be associated with strings $j_1,\dots,j_d$, where $j_i$ corresponds to eigenvalue $\lambda_{j_i}$ of $H$ at position $i$. Assume that we  have chosen $\Delta t$ and $\Delta x$ according to Corollary \ref{cor:accuracy}, such that $\Delta t = \Delta x^2/(2d\alpha)$. Then in order to compute an eigenvalue of $\mathcal{L}$ indexed by $j_1,\dots,j_d$ up to accuracy $\delta'$, it is sufficient to compute each eigenvalue $\lambda_{j_i}$ up to accuracy $O(\delta')$, take the sum, and add 1. Then for the corresponding eigenvalue of $\mathcal{L}^\tau$ to be accurate up to $\delta$, it is sufficient to achieve $\delta'=\delta/\tau$. This follows from all $\mathcal{L}$'s eigenvalues $\lambda$ being in the range $[-1,1]$, which implies that given an approximation $\widetilde{\lambda} = \lambda \pm \delta'$, where $\widetilde{\lambda} \in [-1,1]$, $|\lambda^\tau - \widetilde{\lambda}^\tau| \le \tau \delta'$.

Therefore, we need to compute each eigenvalue $\lambda_j$ up to accuracy $O(\delta/\tau)$. Computing (\ref{eq:lambdaj}) up to $p = O(\log (\tau/\delta))$ digits of precision can be achieved in $O(M(p) \log p)$ time~\cite{brent76}, where $M(p)$ is the complexity of multiplying two $p$-digit integers using some multiplication algorithm (the dominant term in this complexity bound is computing $\sin \theta$). Then raising the sum to the $\tau$'th power can be achieved with additional cost $O(M(p) \log \tau)$.
Choosing $M(p) = O(p^2)$ from standard integer multiplication for simplicity in the final bound, we obtain the stated complexity. 
\end{proof}

\begin{thm}[Classical diagonalisation method]
\label{thm:fftmethod}
There is a classical algorithm that outputs an approximate solution $\widetilde{u}(\mathbf{x},t)$ such that $|\widetilde{u}(\mathbf{x},t) - u(\mathbf{x},t)| \le \epsilon/L^d$ for all $(\mathbf{x},t) \in G$ in time
\be O\left(3^{-d/2} L^d d^{d/2+3} \left(\frac{T\alpha\zeta }{\epsilon}\right)^{d/2} \log^3\left( \frac{TL^2d\alpha\zeta}{\epsilon}\right)\right)
%= \widetilde{O}\left(L^dd^{d/2+3} \left(\frac{T\alpha\zeta}{\epsilon}\right)^{d/2} \right).
\ee
\end{thm}

\begin{proof}
As $\mathcal{L}$ is a sum of circulant matrices acting on $d$ separate dimensions (see (\ref{eq:lform})), it is diagonalised by the $d$-th tensor power of the discrete Fourier transform (equivalently, the inverse quantum Fourier transform up to normalisation). So we use the following expression to approximately compute $\mathbf{\widetilde{u}_i}$:
\be \mathbf{\widetilde{u}_i} = \mathcal{L}^i \mathbf{u_0} = (F^{\otimes d})^{-1} \Lambda^i F^{\otimes d} \mathbf{u_0}, \ee
where $\Lambda$ is the diagonal matrix whose entries are eigenvalues of $\mathcal{L}$, and $F$ is the discrete Fourier transform.
The algorithm begins by writing down $\mathbf{u_0}$ in time $O(n^d)$, then applies the multidimensional fast Fourier transform to $\mathbf{u_0}$ in time $O(d n^d \log n)$ (we assume for simplicity that this step can be performed exactly). Next each entry of the resulting vector is multiplied by the corresponding eigenvalue of $\mathcal{L}^i$, approximately computed up to accuracy $\delta$ using Lemma \ref{lem:computeevs}. Thus we obtain a diagonal matrix $\widetilde{\Lambda^i}$ such that $\|\widetilde{\Lambda^i} - \Lambda^i\| \le \delta$. Then
\be \|(F^{\otimes d})^{-1} \widetilde{\Lambda^i} F^{\otimes d} \mathbf{u_0} -  (F^{\otimes d})^{-1} \Lambda^i F^{\otimes d} \mathbf{u_0} \|_2 \le \|\widetilde{\Lambda^i} -\Lambda^i\|\|\mathbf{u_0}\|_2\le \delta  \|\mathbf{u_0}\|_1 = \delta \left(\frac{n}{L}\right)^d. \ee
So it is sufficient to take $\delta = \epsilon/n^d$. By Lemma \ref{lem:computeevs}, the complexity of the second step is
\be \label{eq:secondstep} O(n^d \log^2(m n^d/\epsilon)(\log \log(mn^d/\epsilon) + \log m)). \ee
The final step of the algorithm is to perform the fast inverse Fourier transform, with equal complexity to the first step, so the second step dominates the overall complexity of the algorithm. 

Choosing $m$ and $n$ according to Corollary \ref{cor:accuracy} gives $m = 2T^2 d^2 \alpha^2  \zeta / (3\epsilon)$, $n = L \sqrt{d \alpha \zeta T/ (3\epsilon)}$. Assuming that $\epsilon \rightarrow 0$ significantly more quickly than $T^2 d^2 \alpha^2 \zeta$ and $L^2 d \alpha \zeta T$ increase, $\log m \gg \log \log(m n^d/\epsilon)$. Using
\be \log(mn^d/\epsilon) = \log\left(3^{-d/2} (Td\alpha/\epsilon)^{d/2+2}\zeta^{d/2+1} L^d \right) = O(d \log(TL^2d\alpha\zeta/\epsilon)), \ee
the overall complexity is
\be O\left(3^{-d/2} L^d d^{d/2+3} \left(\frac{T\alpha\zeta }{\epsilon}\right)^{d/2} \log^3\left( \frac{TL^2d\alpha\zeta}{\epsilon}\right)\right) \ee
as claimed, where we simplify logarithms, bearing in mind dimensions.
\end{proof}

Given a solution that is accurate up to $\ell_\infty$ error $\epsilon/L^d$ at all points in $G$ via Theorem \ref{thm:lineqmethod}, \ref{thm:simplelineqmethod} or \ref{thm:fftmethod}, we can apply Simpson's rule to achieve final error $\epsilon$ in computing the amount of heat in any desired region via numerical integration. This does not increase the overall complexity of any of the above algorithms, as it requires time only $O(n^d)$.

We see that, of all the ``direct'' methods for producing a solution to the heat equation classically, the most efficient is the fast Fourier transform method, which has complexity $\widetilde{O}(3^{-d/2} L^d d^{d/2+3} (T\alpha\zeta/\epsilon)^{d/2})$. However, this only gives us the solution at a particular time $t$, and assumes that we are solving the heat equation in a (hyper)rectangular region.

%-------------------------------------------------------------------------------

\subsection{Random walk method}
\label{sec:randomwalk}

The random walk method for solving the heat equation~\cite{lawler10,kac47,king51} is based around the observation that the linear operator $\mathcal{L}$ corresponding to evolving in time by one step is stochastic, so this process can be understood as a random walk. Given a sample from a distribution corresponding to the initial condition $\mathbf{u}_0$, one can iterate the random walk $m$ times to produce samples from distributions corresponding to each of the subsequent time steps.

\begin{lem}
\label{lem:rwmethod}
Assume that we have chosen particular values for $m$ and $n$. Then there is a classical algorithm that outputs samples from distributions $\overline{\mathbf{u}}_{\mathbf{i}}$ such that $\|\overline{\mathbf{u}}_i - (\Delta x)^d \mathbf{\widetilde{u}_i}\|_\infty \le \epsilon$ for all $i = 0,\dots,m$ in time $O(md\log n)$.
\end{lem}

\begin{proof}
Let $\overline{\mathbf{u}}_{\mathbf{0}} = (\Delta x)^d \mathbf{u_0}$. As $\sum_{(\mathbf{x},0) \in G} u_0(\mathbf{x}) = (\Delta x)^{-d}$, $\overline{\mathbf{u}}_{\mathbf{0}}$ is indeed a probability distribution. We have assumed that $\sum_{(\mathbf{x},0) \in S} u_0(\mathbf{x})$ can be computed without cost, which implies that arbitrary marginals of $\overline{\mathbf{u}}_{\mathbf{0}}$ can be computed without cost. This allows us to sample from $\overline{\mathbf{u}}_{\mathbf{0}}$ in time $O(\log (n^d)) = O(d \log n)$ by a standard technique: split the domain into half and compute the total probability in each region; choose a region to split further, according to these probabilities; and repeat until the region is reduced to just one point $\mathbf{x}$, which is a sample from $\overline{\mathbf{u}}_{\mathbf{0}}$.

Given a sample $\mathbf{x}$ from $\overline{\mathbf{u}}_{\mathbf{i}}$, we can sample from $\overline{\mathbf{u}}_{\mathbf{i+1}} = (\Delta x)^d \mathbf{\widetilde{u}_{i+1}}$ by applying the stochastic map $\mathcal{L}$ to $\mathbf{x}$ (in the sense of sampling from a distribution on new positions, rather than maintaining the entire vector), to update to a new position in time $O(d \log n)$. So we can output one sample from each of the distributions $\overline{\mathbf{u}}_{\mathbf{i}}$ in total time $O(md\log n)$.% = O(Td^3\alpha^2 (\zeta/\epsilon)\log(Ld\alpha\zeta/\epsilon))$.
\end{proof}

We can now use this to approximate the total amount of heat in a given rectangular region at a given time $t$, via the midpoint rule.

\begin{thm}
\label{thm:rwmethodest}
For any $S \subseteq [0,L]^d$ such that the corners of $S$ are all integer multiples of $\Delta x$, shifted by $\Delta x / 2$, and any $t \in [0,T]$ that is an integer multiple of $\Delta t$, there is a classical algorithm that outputs $\overline{u}(S)$ such that $|\overline{u}(S) - \int_S u(\mathbf{x},t) d\mathbf{x}| \le \epsilon$, with probability 0.99, in time
\be O((T\alpha d^3 \zeta(\alpha T + L^2)/\epsilon^3)\log(L \sqrt{d\zeta(\alpha T + L^2)/\epsilon})). \ee
\end{thm}

\begin{proof}
For any probability distribution $P$ and any subset $U$, $\sum_{\mathbf{x} \in U} P(\mathbf{x})$ can be estimated by choosing a sequence of $k$ samples $\mathbf{x}_i$ according to $P$, and outputting the fraction of samples that are contained within $U$. The expectation of this quantity is precisely $\sum_{\mathbf{x} \in U} P(\mathbf{x})$, and by a standard Chernoff bound (or Chebyshev inequality) argument~\cite{dubhashi09}, it is sufficient to take $k = O(1/\epsilon^2)$ to estimate this expectation up to accuracy $\epsilon$ with 99\% probability of success.
We use Lemma \ref{lem:rwmethod} to sample from the required distribution. Let $S'$ denote the set $G \cap S \cap \{(\mathbf{x},t):\mathbf{x} \in [0,L]^d\}$, and write $t = i \Delta t$ for some integer $i$. Then, if we choose $m = O(T\alpha d^2 \zeta(\alpha T + L^2)/\epsilon)$, $n = O(L \sqrt{d\zeta(\alpha T + L^2)/\epsilon})$ (see (\ref{eq:midpointmn})) and apply this technique to $S'$, we get precisely the midpoint rule formula for approximating $\int_S u(\mathbf{x},t) d\mathbf{x}$. Thus we have
\be \left| \sum_{(\mathbf{x},t) \in S'} \overline{\mathbf{u}}_{\mathbf{i}}(\mathbf{x}) - \int_S u(\mathbf{x},t) d\mathbf{x}\right| = O(\epsilon) \ee
via the analysis of the midpoint rule in Section \ref{sec:numint}, noting that we have the normalisation $\overline{\mathbf{u}}_{\mathbf{i}} = (\Delta x)^d \mathbf{\widetilde{u}_i}$. Inserting these choices for $m$ and $n$ into the bound of Lemma \ref{lem:rwmethod} and multiplying by $O(1/\epsilon^2)$ gives the claimed result.
\end{proof}

The reader may wonder why we did not use a differently weighted sum in Theorem \ref{thm:rwmethodest}, corresponding to approximating the integral via Simpson's rule, given that this rule apparently has better accuracy. The reason is that the weighting used for Simpson's rule has components which are exponentially large in $d$, which would lead to an exponential dependence on $d$ in the final complexity, coming from the Chernoff bound.

%-------------------------------------------------------------------------------

\subsection{Fast random walk method}
\label{sec:fastrwmethod}

We can speed up the algorithm of the previous section by sampling from the final distribution of the random walk more efficiently than the na\"ive simulation method of Lemma \ref{lem:rwmethod}.

\begin{lem}
\label{lem:fastrwmethod}
Assume that we have chosen particular values for $m$ and $n$. Then there is a classical algorithm that outputs samples from a distribution $\overline{\mathbf{u}}_{\mathbf{m}}$ such that $\|\overline{\mathbf{u}}_m - (\Delta x)^d \mathbf{\widetilde{u}_m}\|_\infty \le \epsilon$ in expected time $O(d(\log n + \log m))$.
\end{lem}

\begin{proof}
As in Lemma \ref{lem:rwmethod}, we begin by sampling from $\overline{\mathbf{u}}_{\mathbf{0}}$ in time $O(d \log n)$. Next, given such a sample, we want to perform $m$ steps of a random walk on $\Z_n^d$. We can do this by simulating $m$ steps of a random walk on $\Z^d$ and reducing each element of the output modulo $n$. This random walk can be understood as follows: for each of $m$ steps, choose a dimension uniformly at random, then increment or decrement the corresponding coordinate with equal probability of each. The number of steps taken in each dimension can be determined sequentially. For the $i$'th dimension ($1 \le i \le d$), if $m'$ steps have been taken in total in the previous $i-1$ dimensions, the number of steps taken in that dimension is distributed according to a binomial distribution with parameters $(m-m',1/(d-i+1))$. Once the number $s_i$ of steps taken in each dimension $i$ is known, the number of increments in that dimension is also binomially distributed with parameters $(s_i,1/2)$. So the problem reduces to sampling from binomial distributions with parameters $(l,p)$ for arbitrary $l \le m$, $0<p<1$. This can be achieved by combining algorithms described in~\cite[Appendix A.2]{bringmann14} and~\cite[Theorem 2]{farachcolton15}, which allows exact sampling from a binomial distribution using $O(\log m)$ samples from a uniform distribution (in expectation), and expected time $O(\log m)$ (in the ``word RAM'' model which assumes that operations can be performed on $O(\log m)$ bits in constant time). See also~\cite{kachitvichyanukul88,devroye86} for constant-time sampling algorithms, in a model where we assume that operations on real numbers can be performed in constant time.
\end{proof}

We can plug Lemma \ref{lem:fastrwmethod} into the argument of Theorem \ref{thm:rwmethodest} to obtain the following improved result:

\begin{thm}
\label{thm:fastrwmethodest}
For any $S \subseteq [0,L]^d$ such that the corners of $S$ are all integer multiples of $\Delta x$, shifted by $\Delta x / 2$, and any $t \in [0,T]$ that is an integer multiple of $\Delta t$, there is a classical algorithm that outputs $\overline{u}(S)$ such that $|\overline{u}(S) - \int_S u(\mathbf{x},t) d\mathbf{x}| \le \epsilon$, with probability 0.99, in time
\be O((d/\epsilon^2) \log(TL\alpha d^{5/2} \zeta^{3/2} ((\alpha T + L^2)/\epsilon)^{3/2} )). \ee
\end{thm}

\begin{proof}
The proof is the same as for Theorem \ref{thm:rwmethodest}, substituting the use of Lemma \ref{lem:fastrwmethod} for Lemma \ref{lem:rwmethod}. The final complexity is $O(d(\log n + \log m)/\epsilon^2) = O(d(\log nm)/\epsilon^2)$, with $m = O(T\alpha d^2 \zeta(\alpha T + L^2)/\epsilon)$, $n = O(L \sqrt{d\zeta(\alpha T + L^2)/\epsilon})$.
\end{proof}

%-------------------------------------------------------------------------------

\section{Quantum methods}

In this section we describe several quantum algorithms for solving the heat equation. We begin by stating some technical ingredients that we will require.

First, we describe a technical lemma that allows us to go from a quantum state corresponding to an approximate solution to the heat equation at one or more given times simultaneously, to an estimate of the heat in a given region. 

\begin{lem}[Quantum numerical integration]
\label{lem:quantumni}
Let $\mathbf{\widetilde{u}}$ be the $mn^d$-component vector corresponding to some function $\widetilde{u}(\mathbf{x},t)$ such that $|\widetilde{u}(\mathbf{x},t)-u(\mathbf{x},t)| \le \epsilon/L^d$ for all $(\mathbf{x},t) \in G$, and let
\be \ket{\widetilde{u}} = \frac{1}{\sqrt{\sum_{(\mathbf{x},t) \in G} \widetilde{u}(\mathbf{x},t)^2 }} \sum_{(\mathbf{x},t) \in G} \widetilde{u}(\mathbf{x},t) \ket{\mathbf{x},t}, \ee
be the corresponding normalised quantum state. Let $\ket{\widetilde{\widetilde{u}}}$ be a normalised state that satisfies $\|\ket{\widetilde{\widetilde{u}}} - \ket{\widetilde{u}}\|_2 \le \gamma$, where $\gamma = O(\epsilon n^{d/2}/((\sqrt{10}L/3)^d \|\mathbf{\widetilde{u}}\|_2))$. Also assume that we have an estimate $\widetilde{\|\mathbf{\widetilde{u}}\|_2}$ such that $|\widetilde{\|\mathbf{\widetilde{u}}\|_2}-\|\mathbf{\widetilde{u}}\|_2| \le \gamma \|\mathbf{\widetilde{u}}\|_2$. Let $S$ be a hyperrectangular region at a fixed time $t$ such that the corners of $S$ are in $G$. Then it is sufficient to use an algorithm that produces  $\ket{\widetilde{\widetilde{u}}}$ $k$ times to estimate $\int_S u(\mathbf{x},t) d\mathbf{x} \pm \epsilon$, where $k = O( (\sqrt{10}L/3)^d \|\mathbf{\widetilde{u}}\|_2 /( \epsilon n^{d/2}))$.
\end{lem}

\begin{proof}
Let $w(\mathbf{x})$ be a set of weights corresponding to a numerical integration rule as defined in Section \ref{sec:numint} (we will use Simpson's rule in what follows). We will attempt to estimate $\int_S u(\mathbf{x},t) d\mathbf{x}$ by approximately computing $(\Delta x)^d \sum_{\mathbf{x} \in G \cap S} w(\mathbf{x})\widetilde{\|\mathbf{\widetilde{u}}\|_2} \braket{\mathbf{x},t|\widetilde{\widetilde{u}}}$. We first determine the level of accuracy that is required in computing $\widetilde{\|\mathbf{\widetilde{u}}\|_2}$, $\ket{\widetilde{\widetilde{u}}}$.
By the triangle inequality we have
\bea
&& \left| (\Delta x)^d \sum_{\mathbf{x} \in G \cap S} w(\mathbf{x})\widetilde{\|\mathbf{\widetilde{u}}\|_2} \braket{\mathbf{x},t|\widetilde{\widetilde{u}}} - \int_S u(\mathbf{x},t) d\mathbf{x} \right|\\
&\le& (\Delta x)^d \left| \sum_{\mathbf{x} \in G \cap S} w(\mathbf{x}) \widetilde{\|\mathbf{\widetilde{u}}\|_2}\braket{\mathbf{x},t|\widetilde{\widetilde{u}}} - \sum_{\mathbf{x} \in G\cap S} w(\mathbf{x}) \|\mathbf{\widetilde{u}}\|_2 \braket{\mathbf{x},t|\widetilde{\widetilde{u}}} \right|\\
&+& (\Delta x)^d \left| \sum_{\mathbf{x} \in G \cap S} w(\mathbf{x}) \|\mathbf{\widetilde{u}}\|_2\braket{\mathbf{x},t|\widetilde{\widetilde{u}}} - \sum_{\mathbf{x} \in G\cap S} w(\mathbf{x}) \|\mathbf{\widetilde{u}}\|_2 \braket{\mathbf{x},t|\widetilde{u}} \right|\\
&+& \left|(\Delta x)^d \sum_{\mathbf{x} \in G\cap S} w(\mathbf{x}) \|\mathbf{\widetilde{u}}\|_2 \braket{\mathbf{x},t|\widetilde{u}} - \int_S u(\mathbf{x},t) d\mathbf{x} \right|\\
&\le& (\Delta x)^d \left|  \widetilde{\|\mathbf{\widetilde{u}}\|_2} - \|\mathbf{\widetilde{u}}\|_2 \right| \sum_{\mathbf{x} \in G \cap S} \left| w(\mathbf{x}) \braket{\mathbf{x},t|\widetilde{\widetilde{u}}} \right|\\
&+& (\Delta x)^d \|\mathbf{\widetilde{u}}\|_2 \left| \sum_{\mathbf{x} \in G \cap S} w(\mathbf{x}) (\braket{\mathbf{x},t|\widetilde{\widetilde{u}}} - \braket{\mathbf{x},t|\widetilde{u}}) \right|\\
&+& \left|(\Delta x)^d \sum_{\mathbf{x} \in G\cap S} w(\mathbf{x}) \widetilde{u}(\mathbf{x},t) - \int_S u(\mathbf{x},t) d\mathbf{x} \right|\\
&\le& (\Delta x)^d \gamma \|\mathbf{\widetilde{u}}\|_2 \|w\|_2 + (\Delta x)^d \gamma \|\mathbf{\widetilde{u}}\|_2 \|w\|_2 + O(\epsilon)
\eea
%
%where we use $\|w\|_2 = O(\epsilon^{d/4 + 1/2})$.
where in the last inequality we use the analysis of Section \ref{sec:numint} and Cauchy-Schwarz.

To achieve a final bound of $\epsilon$, we need to have $\gamma = O(\epsilon/(\|\mathbf{\widetilde{u}}\|_2(\Delta x)^d  \|w\|_2))$. To find a concrete expression for this requirement, we need to compute $\|w\|_2$. In the case of Simpson's rule, we have
\bea \|w\|_2 &\le& \left(\frac{2}{9} + \frac{n-1}{2} \left(\frac{4}{3}\right)^2 + \frac{n-1}{2} \left(\frac{2}{3}\right)^2 \right)^{d/2}\\
&=& \left(\frac{2}{9} + (n-1)\frac{10}{9} \right)^{d/2} = O((\sqrt{10}/3)^d n^{d/2}). \eea
Thus it is sufficient to take $\gamma = O(\epsilon n^{d/2}/(\sqrt{10}L/3)^d) \|\mathbf{\widetilde{u}}\|_2^{-1}$ to achieve final accuracy $\epsilon$.

Finally, we need to approximately compute $(\Delta x)^d \sum_{\mathbf{x} \in G \cap S} w(\mathbf{x})\widetilde{\|\mathbf{\widetilde{u}}\|_2} \braket{\mathbf{x},t|\widetilde{\widetilde{u}}}$ given an algorithm that produces copies of $\ket{\widetilde{\widetilde{u}}}$. This can be achieved using amplitude estimation~\cite{brassard02} to estimate the inner product between the state
\be \frac{1}{\|w\|_2} \sum_{\mathbf{x} \in G \cap S} w(\mathbf{x}) \ket{\mathbf{x},t} \ee
and $\ket{\widetilde{\widetilde{u}}}$, up to accuracy $\epsilon / ((\Delta x)^d \|w\|_2 \widetilde{\|\mathbf{\widetilde{u}}\|_2})$, and multiplying by $(\Delta x)^d \|w\|_2 \widetilde{\|\mathbf{\widetilde{u}}\|_2}$.
In order to achieve this level of accuracy, we need to use the algorithm for producing $\ket{\widetilde{\widetilde{u}}}$ $k$ times, where $k = O( (\Delta x)^d \|w\|_2 \widetilde{\|\mathbf{\widetilde{u}}\|}/\epsilon)$ from amplitude estimation. Applying the previous calculation of $\|w\|_2$, and using that $\widetilde{\|\mathbf{\widetilde{u}}\|_2} \approx \|\mathbf{\widetilde{u}}\|_2$, gives the claimed result.
\end{proof}

Observe that in fact Lemma \ref{lem:quantumni} can be used to estimate $\int_S u(\mathbf{x},t) d\mathbf{x}$ given copies of states $\ket{\widetilde{\widetilde{u}}}$ corresponding to an approximation to $u$ which is accurate only within $G \cap S$, rather than over all of $S$. We will use this later on to estimate the amount of heat in a region, given a state corresponding to a solution to the heat equation at a particular time $t$, rather than all times as stated in this lemma.

The midpoint rule could be used instead of Simpson's rule in Lemma \ref{lem:quantumni} to integrate over hyperrectangular regions $S$ such that the corners of $S$ are in $G$, shifted by $\Delta x / 2$; this would lead to a similar complexity.
% $\|w\|_2 \le n^{-d/2} = O((\epsilon/(L^5 d\alpha \zeta))^{d/4})$.

We will also need a technical result regarding the $\ell_2$ norm of solutions to the heat equation.

\begin{lem}
\label{lem:l2bound}
Let $\mathcal{L}$ be defined by (\ref{eq:linstep}), taking $\Delta t = \Delta x^2/(2d\alpha)$ as in Corollary \ref{cor:accuracy}. Then for any integer $\tau \ge 1$,
\be \max\left\{\frac{1}{n^d}, \frac{1}{(4\sqrt{\tau})^d} \right\} \le \|\mathcal{L}^\tau\ket{0}\|_2^2 \le d e^{-\tau/(4d)} + \left(\frac{4}{n} + \sqrt{\frac{d}{\pi \tau}}\right)^d. \ee
\end{lem}

In this lemma, and elsewhere, we use $\ket{0}$ to denote the origin in $\R^d$. The proof is deferred to Appendix \ref{app:l2bound}.

%-------------------------------------------------------------------------------

\subsection{Quantum linear equation solving method}
\label{sec:qlinear}

In this section we describe an approach to solve the heat equation using quantum algorithms for linear equations. The idea is analogous to the classical linear equations method: we use a quantum algorithm for solving linear equations to produce a quantum state that encodes a solution approximating $u(\mathbf{x},t)$ for all times $t$, and then use Lemma \ref{lem:quantumni} to estimate  $\int_S u(\mathbf{x},t) d\mathbf{x}$. First we state the complexity of the quantum subroutines that we will use.

\begin{thm}[Solving linear equations~{\cite[Theorem 30 and Corollary 31]{chakraborty18}}]
\label{thm:hhl}
Let $A\mathbf{y}=\mathbf{b}$ for an $N \times N$ matrix $A$ with sparsity $s$ and condition number $\kappa$. Given an algorithm that constructs the state $\ket{b} = \frac{1}{\|\mathbf{b}\|_2}\sum_i \mathbf{b}_i \ket{i}$ in time $T_b$, there is a quantum algorithm that can output a state $\ket{\widetilde{y}}$ such that
\be \left\| \ket{\widetilde{y}} - \ket{y} \right\|_2 \le \eta \ee
in time
\be O\left( \kappa\left( T_U (\log N) \log^2\left(\frac{\kappa}{\eta} \right) + T_b\right) \log \kappa \right), \ee
where
\be T_U = O\left(\log N + \log^{2.5}\left(\frac{s\kappa \log(\kappa/\eta)}{\eta}\right)\right). \ee
\end{thm}

Theorem 30 of \cite{chakraborty18} is stated only for Hermitian matrices, but as remarked in a footnote there, it also applies to non-Hermitian matrices by encoding as a submatrix of a Hermitian matrix. The bound on $T_U$ comes from~\cite[Lemma 48]{gilyen19}. Note that a quantum algorithm by Childs, Kothari and Somma~\cite{childs17b} for solving linear equations could also be used; this would achieve a similar complexity, but the lower-order terms are not stated explicitly in~\cite{childs17b}.

\begin{thm}[Linear equation norm estimation~{\cite[Corollary 32]{chakraborty18}}]
\label{thm:hhlnorm}
Let $A\mathbf{y}=\mathbf{b}$ for an $N \times N$ matrix $A$ with sparsity $s$ and condition number $\kappa$. Given an algorithm that constructs the state $\ket{b} = \frac{1}{\|\mathbf{b}\|_2} \sum_i \mathbf{b}_i \ket{i}$ in time $T_b$, there is a quantum algorithm that outputs $\widetilde{z}$ such that
\be |\widetilde{z} - \|A^{-1}\mathbf{b}\|_2| \le \eta\|A^{-1}\mathbf{b}\|_2 \ee
with probability at least 0.99, in time %$O(s \kappa \poly\log(Ns\kappa/\epsilon))$.
\be O\left(\frac{\kappa}{\eta}\left(T_U (\log N) \log^2\left(\frac{\kappa}{\eta}\right) + T_b \right)(\log^3 \kappa)\log \log\left( \frac{\kappa}{\eta}\right) \right), \ee
where
\be T_U = O\left(\log N + \log^{2.5}\left(\frac{s\kappa \log(\kappa/\eta)}{\eta}\right)\right). \ee
\end{thm}

As the complexity bounds suggest, the algorithms of Theorems \ref{thm:hhl} and \ref{thm:hhlnorm} are rather complicated.

\begin{thm}[Quantum linear equations method]
\label{lem:qprobest}
Let $S \subseteq [0,L]^d$ be a subset at a fixed time $t$. There is a quantum algorithm that produces an estimate $\int_S u(x,t) dx \pm \epsilon$ with 99\% probability of success in time
\be O\left( B L^d 3^{-d/2} (\log^2((Td\alpha)^{d/2+2}(\zeta/\epsilon)^{d/2+1})) (\log^3 ((Td\alpha)^2\zeta/\epsilon)) \log^2 B  \log \log B \right), \ee
where
\be 
B =
\begin{cases} \vspace{.1cm}
O\left(\frac{(T \alpha)^{2.5}\zeta^{1.5}}{\epsilon^{2.5}}(L + \sqrt{T\alpha}) \right) & \text{if }d=1,\\ \vspace{.1cm}
O\left(\frac{(T\alpha)^{2.5} \zeta^{1.5} L}{\epsilon^{2.5}} \sqrt{\log(TL^2\alpha\zeta/\epsilon)} \right) & \text{if }d=2,\\
O\left(\frac{(T\alpha)^{d/4+2}L^{d/2}d^{d/2+2}\zeta^{d/4+1} C^d}{\epsilon^{d/4+2}} \right) & \text{if }d\ge 3,
\end{cases}
\ee
and $C = 20^{1/2}3^{-5/4}\pi^{-1/4}$.
\end{thm}

\begin{proof}
By Corollary \ref{cor:accuracy} and Theorem \ref{thm:condition}, we can achieve discretisation accuracy $\epsilon/L^d$ in the $\infty$-norm with a system of $N=O(mn^d)$ linear equations (see (\ref{linear system:forward method})), each containing $O(d)$ variables, with condition number $\Theta(m)$, where $m = 2 T^2 d^2 \alpha^2  \zeta / (3\epsilon)$, $n = L \sqrt{d \alpha \zeta T/ (3\epsilon)}$. We will apply Theorem \ref{thm:hhl} to solve this system of equations.

First, we can produce the initial quantum state corresponding to the right-hand side of (\ref{linear system:forward method}) as follows. First we construct $\ket{u_0}$, which can be done in time $O(d \log n)$ as we have assumed that we can compute marginals of $u_0$ (and its powers) efficiently~\cite{zalka98,long01,grover02,kaye04}. Then we apply the nonunitary operation $\mathcal{L}$ to $\ket{u_0}$. This can be achieved in time $\widetilde{O}(1/(\kappa d))$, where $\kappa$ is the condition number of $\mathcal{L}$, via an algorithm of~\cite{childs17b}. The $\widetilde{O}$ notation hides polylogarithmic terms in $n^d$. In fact, $\kappa$ can be replaced with $\|\mathcal{L}\|/\|\mathcal{L}\ket{u_0}\|_2$ (see~\cite[Section IIIB]{montanaro16b} for a discussion). From Lemma \ref{lem:conditiononpositive}, and noting that $\|\mathcal{L}\| = O(1)$, this is upper-bounded by $O(\sqrt{d})$. Therefore, the complexity of preparing a normalised version of $\mathcal{L}\ket{u_0}$ is $\poly(d)$ up to logarithmic terms; inspection of Theorem \ref{thm:hhl} shows that this is negligible compared with the complexity of other aspects of the algorithm.

Let $\ket{\widetilde{u}} = \frac{1}{\|\mathbf{\widetilde{u}}\|_2} \sum_{(\mathbf{x},t) \in G} \widetilde{u}(\mathbf{x},t) \ket{\mathbf{x},t}$. Using Theorem \ref{thm:hhl}, there is a quantum algorithm that can produce a state $\ket{\widetilde{\widetilde{u}}}$ such that $\| \ket{\widetilde{\widetilde{u}}} - \ket{\widetilde{u}}\|_2 \le \gamma$ in time
\be \label{lem:qprobest cost to get the state}
O\left( m \log^2 N \log^2 \left(\frac{m}{\gamma}\right) \log m\right) = O\left( m (\log^2 (mn^d)) \log^2 \left(\frac{m}{\gamma}\right) \log m\right). \ee
By Theorem \ref{thm:hhlnorm}, there is a quantum algorithm that produces an estimate $\widetilde{\|\mathbf{\widetilde{u}}\|_2}$ of $\|\mathbf{\widetilde{u}}\|_2$ satisfying
\be 1-\gamma \le \frac{\widetilde{\|\mathbf{\widetilde{u}}\|_2}}{\|\mathbf{\widetilde{u}}\|_2} \le 1+\gamma \ee
in time
\be O\left( \frac{m}{\gamma}(\log^2 (mn^d)) (\log^3 m) (\log^2 \left(\frac{m}{\gamma}\right)) \log \log\left(\frac{m}{\gamma}\right) \right). \ee
In both of these estimates we use that $N \gg \log^{2.5}(dm \log(dm/\gamma)/\gamma)$. Using Lemma \ref{lem:quantumni} and inserting $\gamma = O(\epsilon n^{d/2} /((\sqrt{10}L/3)^d\|\mathbf{\widetilde{u}}\|_2))$, the complexity of producing $\ket{\widetilde{\widetilde{u}}}$ is
\be O\left( m(\log^2 (mn^d))  \log^2 \left(\frac{m(\sqrt{10}L/3)^d \|\mathbf{\widetilde{u}}\|_2}{\epsilon n^{d/2}}\right) \log m\right) \ee
and the complexity of producing $\widetilde{\|\mathbf{\widetilde{u}}\|_2}$ is
\be \label{eq:estimationcomplexity} O\left( \frac{m}{\epsilon n^{d/2}} \left(\frac{\sqrt{10}L}{3} \right)^d \|\mathbf{\widetilde{u}}\|_2 (\log^2 (mn^d)) (\log^3 m) \log^2 \left(\frac{m(\sqrt{10}L/3)^d \|\mathbf{\widetilde{u}}\|_2}{\epsilon n^{d/2}}\right) \log \log\left(\frac{m(\sqrt{10}L/3)^d \|\mathbf{\widetilde{u}}\|_2}{\epsilon n^{d/2}}\right) \right). \ee
By Lemma \ref{lem:quantumni}, in order to estimate $\int_S u(\mathbf{x},t) d\mathbf{x} \pm \epsilon$ it is sufficient to use the algorithm for producing $\ket{\widetilde{\widetilde{u}}}$
\be k = O((\sqrt{10}L/3)^d\|\mathbf{\widetilde{u}}\|_2/(\epsilon n^{d/2})) \ee
times, giving an overall complexity for that part of
\be O\left( \frac{m}{\epsilon n^{d/2}} \left(\frac{\sqrt{10}L}{3} \right)^d\|\mathbf{\widetilde{u}}\|_2 (\log^2(mn^d)) \log^2 \left(\frac{m(\sqrt{10}L/3)^d \|\mathbf{\widetilde{u}}\|_2}{\epsilon}\right)\log m\right). \ee
This implies that the overall complexity of the algorithm is dominated by the complexity of producing the estimate $\widetilde{\|\mathbf{\widetilde{u}}\|_2}$. Defining
\be B = \frac{m}{\epsilon n^{d/2}} \left(\frac{\sqrt{10}L}{3} \right)^d \|\mathbf{\widetilde{u}}\|_2 \ee
for conciseness, (\ref{eq:estimationcomplexity}) can be rewritten as
\be O\left( B (\log^2 (mn^d)) (\log^3 m) \log^2 B  \log \log B \right). \ee
To calculate $B$, it remains to upper-bound $\|\mathbf{\widetilde{u}}\|_2$. A straightforward upper bound is
\be \label{eq:simplel2bound} \|\mathbf{\widetilde{u}}\|_2 = \sqrt{\sum_{i=0}^m \|\mathbf{\widetilde{u}_i}\|_2^2 } \le \sqrt{\sum_{i=0}^{m} \|\mathbf{\widetilde{u}_i}\|_1^2 } = \sqrt{\sum_{i=0}^{m} \left(\frac{n}{L} \right)^{2d}}  = O\left(\sqrt{m} \left(\frac{n}{L} \right)^d \right). \ee
But we will obtain a tighter upper bound, for which it will be sufficient to consider the particular initial condition $\mathbf{u_0}(0^d) = n^d$, $\mathbf{u_0}(\mathbf{x}) = 0$ for $x \neq 0^d$. This initial condition can be seen to give a worst-case upper bound by convexity, as follows. Consider the operator $\mathcal{L}$ occurring in (\ref{eq:linstep}) and an arbitrary initial condition $\mathbf{u'}(\mathbf{x}) = p_{\mathbf{x}}$ such that $\sum_{\mathbf{x}} p_{\mathbf{x}} = (n/L)^d$ (corresponding to the $L_1$ norm of the initial condition being normalised to 1). Then $\mathbf{u_0}$ is a convex combination of point functions of the form $\mathbf{u_{x_0}}(\mathbf{x_0}) = (n/L)^d$, $\mathbf{u_{x_0}}(\mathbf{x}) = 0$ for $\mathbf{x} \neq \mathbf{x_0}$. So $\|\mathcal{L}^\tau \mathbf{u'}\|_2 \le \|\mathcal{L}^\tau\mathbf{u_0}\|_2$ by convexity of the $\ell_2$ norm and shift-invariance of $\mathcal{L}$.

By Lemma \ref{lem:l2bound}, for any $\tau \ge 1$,
\be \|\mathcal{L}^\tau \mathbf{u_0}\|_2^2  \le \left(\frac{n}{L}\right)^{2d} \left( d e^{-\tau/(4d)} + \left(\frac{4}{n} + \sqrt{\frac{d}{\pi \tau}}\right)^d \right). \ee
This gives an upper bound on the total $\ell_2$ norm of
\begin{eqnarray}
\sqrt{\sum_{\tau=0}^{m} \|\mathcal{L}^\tau \mathbf{u_0}\|_2^2 } &\le& \left(\frac{n}{L}\right)^d \sqrt{1 + d \sum_{\tau=1}^m e^{-\tau/(4d)} + \sum_{\tau=1}^m \left(\frac{4}{n} + \sqrt{\frac{d}{\pi \tau}}\right)^d}\\
 &\le& \left(\frac{n}{L}\right)^d \sqrt{1 + d \sum_{\tau \ge 0} e^{-\tau/(4d)} + 2^d \!\!\!\!\sum_{1 \le \tau \le n^2d/(16\pi)} \!\!\!\!\left(\frac{d}{\pi \tau}\right)^{d/2}  + 2^d\!\!\!\! \sum_{n^2d/(16\pi) \le \tau \le m}\!\!\!\! \left(\frac{4}{n}\right)^d }\\
 &\le& \left(\frac{n}{L}\right)^d \sqrt{1 + \frac{d}{1-e^{-1/(4d)}} + \left(\frac{4d}{\pi}\right)^{d/2} \sum_{1 \le \tau \le n^2d/(16\pi)} \tau^{-d/2} + m \left(\frac{8}{n}\right)^d }.
\end{eqnarray}
The first two summands under the square root are negligible compared with the others. For $d=1$, the sum over $\tau$ is $O(n)$; for $d=2$, it is $O(\log n)$; and for $d \ge 3$, it is $O(1)$.  The final summand is negligible for $d \ge 2$ (but not for $d=1$), in the usual situation that $T$, $L$, $d$, $\alpha$, $\epsilon$ and $\zeta$ are such that $m/n^d = O(1)$. This then gives us overall $\ell_2$ norm bounds $\|\mathbf{\widetilde{u}}\|_2 = O((n^{3/2} + \sqrt{mn})/ L)$ for $d=1$, $\|\mathbf{\widetilde{u}}\|_2 = O(n^2 \sqrt{\log n}/L^2)$ for $d=2$, and $\|\mathbf{\widetilde{u}}\|_2 = O((\sqrt{2}d^{1/4}n/(\pi^{1/4} L))^d)$ for $d\ge 3$. Compared with (\ref{eq:simplel2bound}), this last bound is stronger by a factor of almost $\sqrt{m}$. By the lower bound part of Lemma \ref{lem:l2bound}, the bounds are close to tight.

Inserting the values for $m$ and $n$, and these bounds on $\|\mathbf{\widetilde{u}}\|_2$, in the complexity bound (\ref{eq:estimationcomplexity}), the final complexities are as stated in the theorem. In computing these, 
%for conciseness we upper-bound $n^{3/2} + \sqrt{mn}$ by $O(L^{3/2} T^{5/4}\alpha(\zeta/\epsilon)^{3/4})$,
we use the bounds that
\be 
B = \frac{m (\sqrt{10}L/3)^d \|\mathbf{\widetilde{u}}\|_2}{\epsilon n^{d/2}} = 
\begin{cases} \vspace{.1cm}
O\left(\frac{(T\alpha)^{2.5}\zeta^{1.5}}{\epsilon^{2.5}}(L + \sqrt{T\alpha}) \right) & \text{if }d=1,\\ \vspace{.1cm}
O\left(\frac{(T\alpha)^{2.5} \zeta^{1.5} L}{\epsilon^{2.5}} \sqrt{\log(TL^2\alpha\zeta/\epsilon)} \right) & \text{if }d=2,\\
O\left(\frac{(T\alpha)^{d/4+2}L^{d/2}d^{d/2+2}\zeta^{d/4+1} C^d}{\epsilon^{d/4+2}} \right) & \text{if }d\ge 3,
\end{cases}
\ee
where $C = 20^{1/2}3^{-5/4}\pi^{-1/4}$.
\end{proof}

Note that in this analysis, as in the classical case, we have assumed that arbitrary nonzero entries of the matrix $A$ can be computed in time $O(1)$.
% ------------------------------------------------------------------------------

\subsection{Fast-forwarded random walk method}

We next consider alternative methods which directly produce a quantum state corresponding to the distribution of the random walk at time $t = i \Delta t$: that is, a state $\ket{\psi_i}$ close to $\sum_{\mathbf{x}} \widetilde{u}_i(\mathbf{x}) \ket{\mathbf x}/ \|\mathbf{\tilde u_i}\|_2$. We can then estimate $\int_S u(\mathbf{x},t)d\mathbf{x} \pm \epsilon$ using Lemma \ref{lem:quantumni}.

These methods start by producing an initial state $\ket{u_0} = \sum_{\mathbf{x}} u_0(\mathbf{x}) \ket{\mathbf x}/ \|\mathbf{u_0}\|_2$. Given that we have assumed that we can compute sums of squares of $u_0$ over arbitrary regions in time $O(1)$, $\ket{u_0}$ can be constructed in time $O(d \log n)$ via the techniques of~\cite{zalka98,grover02,kaye04}. This will turn out not to affect the overall complexity of the algorithms.

The first approach we consider can be viewed as a coherent version of the random walk method. Given the initial state $\ket{u_0}$, we attempt to produce a state approximating $\ket{u_i} = \ket{\mathcal{L}^i u_i}$ for some $i$.

\begin{thm}[Apers and Sarlette~\cite{apers18}, Gily\'en et al.~\cite{gilyen19}]
\label{thm:ffwalk}
Given a symmetric Markov chain with transition matrix $\mathcal{L}$ and a quantum state $\ket{\psi_0}$, there is an algorithm which produces a state $\ket{\widetilde{\psi_i}}$ such that
\be \left\|\ket{\widetilde{\psi_i}} -  \frac{\mathcal{L}^i\ket{\psi_0}}{\|\mathcal{L}^i \ket{\psi_0}\|_2} \right\|_2 \le \eta \ee
using
\be 
O\left(\|\mathcal{L}^i\ket{\psi_0}\|_2^{-1} \sqrt{i \log(1/(\eta\|\mathcal{L}^i\ket{\psi_0}\|_2))}\right)
\ee
steps of the quantum walk corresponding to $\mathcal{L}$.
\end{thm}

\begin{thm}[Fast-forwarded random walk method]
\label{thm:ffrwmethod}
Let $S$ be a subset at a fixed time $t = i\Delta t$. There is a quantum algorithm based on fast-forwarding random walks that estimates $\int_S u(\mathbf{x},t) d\mathbf{x} \pm \epsilon$ in time
\be O\left(\frac{d^{5/2}T\alpha\zeta^{1/2}}{\epsilon^{3/2}}\
\left(\frac{100 L^2 d T\alpha\zeta}{3^5\epsilon}\right)^{d/4}
\log(L^2d\alpha\zeta T/\epsilon) \sqrt{\log(L^2d\alpha\zeta T/\epsilon)} \right). \ee
\end{thm}

\begin{proof}
We use the algorithm of Theorem \ref{thm:ffwalk} to produce a state $\ket{\widetilde{\widetilde{u}}_i}$ such that $\| \ket{\widetilde{\widetilde{u}}_i} - \ket{\widetilde{u}_i} \|_2 \le \gamma$, where $\ket{\widetilde{u}_i} = \mathbf{\widetilde{u}_i}/\|\mathbf{\widetilde{u}_i}\|_2$ and $\gamma$ is defined in Lemma \ref{lem:quantumni}, which is applied at a single time. We need to use this algorithm $k$ times, where $k$ is also defined in Lemma \ref{lem:quantumni}. The complexity of implementing a quantum walk step is essentially the same as that of implementing a classical random walk step, which is $O(d \log n)$. The complexity of producing the initial state $\ket{u_0}$ is also $O(d \log n)$. Therefore, the complexity of the overall algorithm is
\be 
O\left(
d (\log n) k \|\mathbf{u_0}\|_2 \|\mathbf{\widetilde{u}_i}\|_2^{-1} \sqrt{m \log(\|\mathbf{u_0}\|_2/(\gamma\|\mathbf{\widetilde{u}_i}\|_2))}
\right). 
\ee
As $k = O( (\sqrt{10}L/3)^d \|\mathbf{\widetilde{u}_i}\|_2 / (\epsilon n^{d/2}))$, $\gamma = O(\epsilon n^{d/2}/((\sqrt{10}L/3)^d \|\mathbf{\widetilde{u}_i}\|_2))$ from Lemma \ref{lem:quantumni}, we see that the $\|\mathbf{\widetilde{u}_i}\|_2$ terms cancel.
Inserting the values for $\gamma$ and $k$, using $\|\mathbf{u_0}\|_2 \le (n/L)^d$ and inserting the values for $n$ and $m$ determined in Corollary \ref{cor:accuracy}, we obtain the claimed result.
\end{proof}

% ------------------------------------------------------------------------------

\subsection{Diagonalisation and postselection method}
\label{sec:diag}

Similarly to the classical case (Theorem \ref{thm:fftmethod}), we can find a more efficient algorithm than Theorem \ref{thm:ffrwmethod} (one without the factor of $\sqrt{m}$) in the special case we are considering of solving the heat equation in a hypercube, using the fact that the quantum Fourier transform diagonalises $\mathcal{L}$. By contrast with the classical method, here we perform operations in superposition. As in the previous section, again the goal is to produce $\ket{u_i}$ for some $i$; as we can diagonalise $\mathcal{L}$ efficiently, all that remains is to implement the (non-unitary) operation $\Lambda^i$, where $\Lambda$ is the diagonal matrix corresponding to the eigenvalues of $\mathcal{L}$.

\begin{thm}[Quantum diagonalisation and postselection method]
\label{thm:diag}
Let $S$ be a hyperrectangular region at a fixed time $t = i\Delta t$ such that the corners of $S$ are in $G$. There is a quantum algorithm that estimates $\int_S u(\mathbf{x},t) d\mathbf{x} \pm \epsilon$ with 99\% sucess probability in time
\be O\left(\left(\frac{100L^2 dT\alpha\zeta}{3^5} \right)^{d/4} \epsilon^{-d/4-1} (\log^2 D)(\log \log D + \log(T^2 d^2\alpha^2\zeta/\epsilon))\right), \ee
where
\be D = O\left( \left(\frac{\sqrt{10L}}{3^{5/4}} \right)^d \frac{(Td\alpha)^{d/4+2}\zeta^{d/4+1}}{\epsilon^{d/4+2}} \right). \ee
\end{thm}

\begin{proof}
We start with the state $\ket{u_0}$, and apply the approximate quantum Fourier transform in time $O(d \log n \log \log n)$ to produce a state $\ket{\psi}$. Note that this is exponentially faster than the classical FFT. Then, similarly to Theorem \ref{thm:fftmethod}, we want to apply the map $\Lambda^i$ to this state, where $\Lambda$ is the diagonal matrix whose entries are eigenvalues of $\mathcal{L}$, before applying the inverse quantum Fourier transform to produce $\ket{\widetilde{u}_i}$. Recalling that eigenvalues $\lambda_j$ of $\mathcal{L}$ correspond to strings $j = j_1,\dots,j_d$, where $j_1,\dots,j_d \in \{0,\dots,n-1\}$, we expand
\be \ket{\psi} = \sum_{j_1,\dots,j_d=0}^{n-1} \psi_{j_1,\dots,j_d} \ket{j_1,\dots, j_d}. \ee
Then applying $\Lambda^i$ can be achieved by performing the map
\be \label{eq:ancillamap} \ket{\psi}\ket{0} \mapsto \sum_{j_1,\dots,j_d=0}^{n-1} \psi_{j_1,\dots,j_d} \ket{j_1,\dots,j_d}\left(\lambda^i_j\ket{0} + \sqrt{1-\lambda_j^{2i}}\ket{1}\right) \ee
and measuring the ancilla qubit. If we receive the outcome 0, then the residual state is as desired, and we can apply the inverse quantum Fourier transform to produce $\mathcal{L}^i \ket{u_0}/\|\mathcal{L}^i \ket{u_0}\|_2$. The probability that the measurement of the ancilla qubit succeeds is precisely $\|\mathcal{L}^i \ket{u_0}\|_2^2$. Using amplitude amplification, $O(\|\mathcal{L}^i \ket{u_0}\|_2^{-1})$ repetitions are enough to produce the desired state with success probability 0.99. We will also need to produce an estimate of $\|\mathbf{\widetilde{u}_i}\|_2$. To do so, we can apply amplitude estimation to this procedure to produce an estimate of the square root of the probability of receiving outcome 0. This gives $\|\mathbf{\widetilde{u}_i}\|_2 (1\pm\delta)$ (with success probability lower-bounded by a constant arbitrarily close to 1) at an additional multiplicative cost of $O(\delta^{-1})$~\cite{brassard02}.

For any $i \in \{0,\dots,m\}$, and any $\delta > 0$, by Lemma \ref{lem:computeevs} each eigenvalue of $\mathcal{L}^i$ can be computed classically up to accuracy $\delta$ in time $O(\log^2 (m/\delta)(\log \log (m/\delta) + \log m))$. Given such an algorithm, we can perform the map (\ref{eq:ancillamap}) on the ancilla qubit up to accuracy $O(\delta)$ as follows. It is shown in~\cite[Section 4.3]{cao13} that to produce the state $\omega \ket{0} + \sqrt{1-\omega^2}\ket{1}$ given knowledge of an approximation to $\omega$, it is sufficient to compute $\theta = \arcsin{\omega}$ and then use $O(\log 1/\delta)$ controlled-Y operations. Computing $\arcsin$ up to $p$ digits of precision can be achieved in $O(M(p) \log p)$ time~\cite{brent76}, where $M(p) = O(p^2)$ is the complexity of multiplying two $p$-digit integers using some multiplication algorithm. Therefore, the additional cost is an additive $O(\log^2(1/\delta)\log \log(1/\delta))$ term, which is negligible.

Thus the overall cost of producing the state $\mathcal{L}^i \ket{u_0}/\|\mathcal{L}^i \ket{u_0}\|_2$ is
\be \label{eq:overallcost} O(\|\mathcal{L}^i \ket{u_0}\|_2^{-1}(d \log n \log \log n + \log^2 (m/\delta)(\log \log (m/\delta) + \log m))). \ee
In order to use Lemma \ref{lem:quantumni}, we need to have $\delta \le \gamma = O(\epsilon n^{d/2}/((\sqrt{10}L/3)^d \|\mathbf{\widetilde{u}_i}\|_2))$. Using this, we get $d \log n = O(\log(m/\delta))$, implying that the $O(d \log n \log \log n)$ term (the cost of implementing the QFT) is negligible.

For this sufficiently small choice of $\delta$, by Lemma \ref{lem:quantumni} we can use the above procedure $k$ times to estimate $\int_S u(\mathbf{x},t) d\mathbf{x} \pm \epsilon$, where $k = O( (\sqrt{10}L/3)^d \|\mathbf{\widetilde{u}_i}\|_2 /(\epsilon n^{d/2})) = O(1/\delta)$. So we see that the complexity of producing a sufficiently accurate estimate of $\|\mathbf{\widetilde{u}_i}\|_2$ is asymptotically equivalent to that of performing the numerical integration. Simplifying (\ref{eq:overallcost}) by using $\ket{u_0} = \frac{1}{\|\mathbf{u_0}\|_2} \sum_{\mathbf{x}} u_0(\mathbf{x}) \ket{\mathbf{x}}$, a $\|\mathbf{\widetilde{u}_i}\|_2$ term cancels, leaving a cost of
\be O(\|\mathbf{u_0}\|_2 (\sqrt{10}L/3)^d \epsilon^{-1} n^{-d/2} \log^2 (m/\delta)(\log \log (m/\delta) + \log m)). \ee
Inserting the values for $m$, $n$ and $\delta$ based on Corollary \ref{cor:accuracy} and using the upper bounds $\|\mathbf{\widetilde{u}_i}\|_2 \le \|\mathbf{u_0}\|_2 \le \|\mathbf{u_0}\|_1 \le (n/L)^d$, we define
\be D = \frac{m}{\delta} = O\left( \left(\frac{\sqrt{10L}}{3^{5/4}} \right)^d \frac{(Td\alpha)^{d/4+2}\zeta^{d/4+1}}{\epsilon^{d/4+2}} \right) \ee
and obtain an overall bound of
\be O\left(\left(\frac{100L^2 dT\alpha\zeta}{3^5} \right)^{d/4} \epsilon^{-d/4-1} (\log^2 D)(\log \log D + \log(T^2 d^2\alpha^2\zeta/\epsilon))\right) \ee
as claimed in the theorem.
\end{proof}

% ------------------------------------------------------------------------------

\subsection{Random walk amplitude estimation approach}
\label{sec:rwamp}

In our final algorithms, we apply amplitude estimation to the classical random walk approach of Section \ref{sec:randomwalk} and Section \ref{sec:fastrwmethod}. This is the simplest of all the quantum approaches, but turns out to achieve the most efficient results in most cases. We begin with the application to accelerating the ``standard'' random walk method.

\begin{thm}
\label{thm:accelrwmethodest}
For any $S \subseteq [0,L]^d$ such that the corners of $S$ are all integer multiples of $\Delta x$, shifted by $\Delta x / 2$, and any $t \in [0,T]$ such that $t = i \Delta t$ for some integer $i$, there is a quantum algorithm that outputs $\overline{u}(S)$ such that $|\overline{u}(S) - \int_S u(\mathbf{x},t) d\mathbf{x}| \le \epsilon$, with probability 0.99, in time $O((T\alpha d^3 \zeta(\alpha T + L^2)/\epsilon^2)\log(L \sqrt{d\zeta(\alpha T + L^2)/\epsilon}))$.
\end{thm}

\begin{proof}
The argument is the same as Theorem \ref{thm:rwmethodest}, except that we use amplitude estimation~\cite{brassard02}, rather than standard probability estimation.
Given a classical boolean function $f$ that takes as input a sequence $s$ of bits, amplitude estimation allows $\Pr_s[f(s)=1]$ to be estimated up to accuracy $\epsilon$, with success probability 0.99, using $f$ $O(1/\epsilon)$ times. In this case, we can think of $s$ as the random seed input to a deterministic procedure which first produces a sample from $\mathbf{\overline{u}}_0$, where $\overline{\mathbf{u}}_{\mathbf{0}} = (\Delta x)^d \mathbf{u_0}$ as in Lemma \ref{lem:rwmethod}, and then executes a sequence of $i$ steps of the random walk. Then $f(s)=1$ if the final position is within $S$, and $f(s)=0$ otherwise. This can be used to estimate $\int_S u(\mathbf{x},t) d\mathbf{x}$ in the same way as the proof of Theorem \ref{thm:rwmethodest}, except that the complexity is lower by a factor of $\Theta(1/\epsilon)$.
\end{proof}

Note that this approach as described in Theorem \ref{thm:accelrwmethodest} uses space $O(m) = O(T^2 d^2 \alpha^2 \zeta/\epsilon)$ to store the sequence of movements of the random walk. This is substantially worse than the classical equivalent, which uses space $O(d \log n) = O(d \log(L^2Td\alpha \zeta/\epsilon))$. It has been an open problem since 2001 whether quantum algorithms can coherently simulate general classical random walk processes with little space overhead~\cite{watrous01}. However, quadratic space overhead over the classical algorithm (which is sufficient to give a polylogarithmic space quantum algorithm) can be achieved using the pseudorandom number generator of Nisan~\cite{nisan92} to replace the sequence of $O(m)$ random bits specifying the movements of the walk.

% ------------------------------------------------------------------------------

\subsection{Fast random walk amplitude estimation approach}
\label{sec:fastrwamp}

Finally, we can also apply amplitude estimation to speed up the algorithm of Theorem \ref{thm:fastrwmethodest}.

\begin{thm}
\label{thm:fastaccelrwmethodest}
For any $S \subseteq [0,L]^d$ such that the corners of $S$ are all integer multiples of $\Delta x$, shifted by $\Delta x / 2$, and any $t \in [0,T]$ such that $t = i \Delta t$ for some integer $i$, there is a quantum algorithm that outputs $\overline{u}(S)$ such that $|\overline{u}(S) - \int_S u(\mathbf{x},t) d\mathbf{x}| \le \epsilon$, with probability 0.99, in time $ O((d/\epsilon) \log(TL\alpha d^{5/2} \zeta^{3/2} ((\alpha T + L^2)/\epsilon)^{3/2} ))$.
\end{thm}

\begin{proof}
The argument is the same as the proof of Theorem \ref{thm:accelrwmethodest}. We apply amplitude amplification to the random seed used as input to a procedure for sampling from the initial distribution and the binomial distributions required for the corresponding classical random walk algorithm (Theorem \ref{thm:fastrwmethodest}). As in the case of Theorem \ref{thm:accelrwmethodest}, the complexity is lower than the corresponding classical algorithm by a factor of $\Theta(1/\epsilon)$.
\end{proof}

%-------------------------------------------------------------------------------

\section{Concluding remarks}

We have considered ten algorithms (five classical and five quantum) for solving the heat equation in a hyperrectangular region, and have found that the quantum algorithm for solving linear equations is never the fastest, but that for $d\ge 2$, a quantum algorithm based on applying amplitude amplification is the most efficient, achieving a speedup up to quadratic over the fastest classical algorithm. However, quantum algorithms based on solving linear equations may have other advantages over the classical ones, such as flexibility for more complicated problems, and better space-efficiency.

The heat equation is of interest in itself, but also as a model for understanding the likely performance of quantum algorithms when applied to other PDEs. For example, it was claimed in~\cite{cao13} that a quantum algorithm for solving Poisson's equation could achieve an exponential speedup over classical algorithms in terms of the spatial dimension $d$. However, Poisson's equation can be solved using a classical random walk method which remains polynomial-time even for large $d$~\cite{bauer58}; this method approximates the solution at a particular point, rather than giving the solution in a whole region. It seems likely that other classical approaches to solving PDEs may be able to compete with some apparent exponential quantum speedups, analogously to the ``dequantization'' approach in quantum machine learning (see~\cite{chia19} and references therein).

%\footnote{The complexity analysis of~\cite{cao13} does not take into account the need for a repeated measurement / postselection step (step vi in the algorithm of Section 4 of~\cite{cao13}), but even taking this into account, the complexity would not increase exponentially with $d$.}

%-------------------------------------------------------------------------------

\subsection*{Acknowledgements}

We would like to thank Jin-Peng Liu and Gui-Lu Long for comments on a previous version. We acknowledge support from the QuantERA ERA-NET Cofund in Quantum Technologies implemented within the European Union's Horizon 2020 Programme (QuantAlgo project), EPSRC grants EP/R043957/1 and EP/T001062/1, and EPSRC Early Career Fellowship EP/L021005/1. This project has received funding from the European Research Council (ERC) under the European Union's Horizon 2020 research and innovation programme (grant agreement No.\ 817581). No new data were created during this study.

%-------------------------------------------------------------------------------

\bibliography{heatequation}

%-------------------------------------------------------------------------------

\appendix

%-------------------------------------------------------------------------------

\section{Runtime of applying a quantum algorithm for ODEs to the heat equation}
\label{app:odealgm}

In this appendix, we sketch the complexity obtained when using the algorithm of Berry et al.~\cite{berry17} to solve the heat equation as a system of ODEs. 
Also note that a procedure is not explicitly given in~\cite{berry17} to approximate the $\ell_2$ norm of the solution vector, which is required to estimate its properties.
We will show that the quantum algorithm based on \cite{berry17}
is somewhat worse than the quantum linear equations method proposed in
Theorem \ref{lem:qprobest} 
to generate the quantum state of the heat equation.

In the heat equation (\ref{eq:heatequation}), if we just discretise
$x_1,\ldots,x_d$ to the same level of accuracy as specified in Section \ref{sec:discretisation}, then we obtain a system of ODEs of the form
\be \label{heat-equation:ODE}
\frac{d \widetilde{\mathbf{u}}}{ dt} = \frac{\alpha}{\Delta x^2} A \widetilde{\mathbf{u}},
\ee
where $\widetilde{\mathbf{u}}$ is the vector of 
$\{u(j_1\Delta x,\ldots,j_d\Delta x,t): ~ j_1,\ldots,j_d\in\{0,1,\ldots,n-1\}\}$,
\be  \label{ODE coeff matrix}
A =  \sum_{j=1}^d I_{n}^{\otimes (j-1)} \otimes H \otimes I_{n}^{\otimes (d-j)},
\ee
and 
\be \label{matrix-H}
H =
\left(
  \begin{array}{cccccc}
    -2 & 1  &        &&1                  \\
    1  & -2 & 1                        \\
       & 1  & \ddots  & \ddots         \\
       &    & \ddots  & \ddots  & 1       \\
     1  &    &    &  1 & -2
  \end{array}
\right)
\ee
is an $n\times n$ matrix.

In \cite{berry17}, Berry et al. proposed a quantum algorithm to
solve time-independent ODEs $\frac{d\mathbf{x}}{dt} = 
A \mathbf{x} + \mathbf{b}$.  They 
assumed that $A$ is diagonalizable and the real 
parts of the eigenvalues are non-positive.
This is satisfied for the heat equation (\ref{heat-equation:ODE})
as shown in the following lemma.

\begin{lem} \label{eigenvalyes of A}
The eigenvalues of $A$ are $\{
\lambda_{j_1}+\cdots+\lambda_{j_d}:
j_1,\ldots,j_d \in\{0,1,\ldots,n-1\}\}$, where
\be
\lambda_j =-4 \sin^2 \frac{j \pi}{n}.
\ee
Moreover, $A$ is diagonalized by the $d$-th tensor product
of the quantum Fourier transform.
\end{lem}

\begin{proof}
Since $H$ is a circulant matrix, it can be diagonalized by the quantum Fourier transform $F$. Denote $\Lambda={\rm diag}\{\lambda_0,\ldots,\lambda_{n-1}\}$ as the diagonal matrix that stores the eigenvalues of $H$, then $\Lambda F^\dag = F^\dag H$.
Set $c_0=-2,c_1=1,c_2=\cdots=c_{n-2}=0,c_{n-1}=1$. Then $\Lambda F^\dag |0\rangle= F^\dag H |0\rangle$
gives
\be
\frac{1}{\sqrt{n}}
\left(
  \begin{array}{c}
    \lambda_0 \\
    \lambda_1 \\
    \vdots \\
    \lambda_{n-1} \\
  \end{array}
\right) = F^\dag
\left(
  \begin{array}{c}
    c_0 \\
    c_1 \\
    \vdots \\
    c_{n-1} \\
  \end{array}
\right).
\ee
For convenience, set $\omega_{n}=e^{2\pi i/n}$, then
\be \label{eigenvalues of H}
\lambda_j = \sum_{k=0}^{n-1} c_k \omega_{n}^{-jk}
=-2+ \omega_{n}^{-j}+\omega_{n}^{-j(n-1)}
=-2+ \omega_{n}^{-j}+\omega_{n}^{j}
=-2+2 \cos \frac{2j \pi}{n}
=-4 \sin^2 \frac{j \pi}{n}.
\ee
The claimed result follows easily from
equation (\ref{ODE coeff matrix}).
\end{proof}

Since we can determine the nonzero entries of $H$ efficiently,
we can determine the nonzero entries of $A$ efficiently too. The sparsity of 
$A$ is $\Theta(d)$.
By Theorem 9 of \cite{berry17}, 
the quantum state $|\widetilde{\mathbf{u}}(T)\rangle$ of the ODE 
(\ref{heat-equation:ODE}) to precision $\epsilon$ is obtained in time
\be
\widetilde{O}(dgT\|A\|),
\ee
where 
$g=\max_{t\in[0,T]} \|\widetilde{\mathbf{u}}(t)\|/\|\widetilde{\mathbf{u}}(T)\|$.
By Lemma \ref{eigenvalyes of A} and Corollary \ref{cor:accuracy}, 
\be
\|A\| = \frac{\alpha}{\Delta x^2} \max_{j_1,\ldots,j_d}
|\lambda_{j_1}+\cdots+\lambda_{j_d}|
=\frac{4\alpha d}{\Delta x^2}
=\Theta(\alpha^2 d^2\zeta T/\epsilon).
\ee
Thus, the quantum state $|\widetilde{\mathbf{u}}(T)\rangle$ is obtained
in time
\be \label{complexty-berry}
\widetilde{O}(\alpha^2 d^3 T^2 g \zeta /\epsilon).
\ee

Note that in the proof of Theorem \ref{lem:qprobest},
equation (\ref{lem:qprobest cost to get the state}) shows that
we can obtain the state $|\widetilde{u}\rangle$ in time
\be
\widetilde{O}(md^2)=\widetilde{O}(\alpha^2 d^4T^2  \zeta /\epsilon).
\ee
In comparison, the complexity of the algorithm of \cite{berry17} has better dependence on $d$, but
is increased by a multiplicative factor $g \ge 1$. The complexity of obtaining the desired state
using the quantum spectral method of 
Childs and Liu~\cite{childs20} also
equals (\ref{complexty-berry}).

%-------------------------------------------------------------------------------

\section{Estimation of the condition number}
\label{appendix:Estimation of the condition number}

For some of the classical and quantum methods we consider, the condition number of the relevant linear system will be an important component of the algorithms' overall complexity.

Recall from equation (\ref{linear system:forward method}) that this linear system is
\be \label{appendix-linear system:forward method}
\left(
  \begin{array}{cccccc}
    I &   &        &&     \\
    -\mathcal{L}  & I           \\
       &     \ddots  & \ddots  &     \\
      &        &  -\mathcal{L} & I
  \end{array}
\right)
\left(
  \begin{array}{c}
    \mathbf{\widetilde{u}_1}     \\
    \mathbf{\widetilde{u}_2}           \\
    \vdots     \\
    \mathbf{\widetilde{u}_m}
  \end{array}
\right)
=\left(
  \begin{array}{c}
    \mathcal{L}\mathbf{\widetilde{u}_0}     \\
    0           \\
    \vdots     \\
    0
  \end{array}
\right).
\ee
In the following, we will estimate the condition number of the above linear system. For convenience, we let $A$ denote the coefficient matrix.

First we consider the case $d=1$.
% Denote 
% \be 
% H =
% \left(
%   \begin{array}{cccccc}
%     -2 & 1  &        &&1                  \\
%     1  & -2 & 1                        \\
%       & 1  & \ddots  & \ddots         \\
%       &    & \ddots  & \ddots  & 1       \\
%      1  &    &    &  1 & -2
%   \end{array}
% \right),
% \ee
% which is an $(n+1)\times (n+1)$ matrix,
In this case
\be
\mathcal{L} = I + \frac{\alpha \Delta t}{\Delta x^2} H,
\ee
where $H$ is the matrix defined in equation (\ref{matrix-H}).
If we define $\mathcal{T}$ to be the following $m\times m$ matrix:
\be \label{matrixT}
\mathcal{T} = \left(
      \begin{array}{cccccc}
         1  &  \\
            -1 & 1 \\
               & \ddots & \ddots \\
               &        & -1 & 1
      \end{array}
    \right),
\ee
then
\be
A = \mathcal{T} \otimes \mathcal{L} - \frac{\alpha \Delta t}{\Delta x^2} I \otimes H.
\ee

For convenience, denote
\be \label{gamma}
\gamma_j = 4\frac{\alpha \Delta t}{\Delta x^2}\sin^2 \frac{j \pi}{n}, ~~~ j=0,1,\ldots,n-1.
\ee
Then by Lemma \ref{eigenvalyes of A},
the eigenvalues of $\mathcal{L}$ are $1-\gamma_j$ for $j=0,1,\ldots,n-1$.
Moreover,
\be
(I\otimes F^\dag) A (I\otimes F)
= \sum_{j=0}^{n-1} \left((1-\gamma_j) \mathcal{T} + \gamma_j I\right)
\otimes |j\rangle \langle j|,
\ee
where $F$ is the quantum Fourier transform. It is easy to show that the set of singular values of $A$ is the collection of the singular values of
\be
A_j = (1-\gamma_j) \mathcal{T} + \gamma_j I
\ee
for all $j$. 
Next, we focus on the calculation of
the singular values of $A_j$. Note that if $\gamma_j=1$, then $A_j=I$. This case is trivial, so we assume that $\gamma_j\neq1$ in the following. From equation
(\ref{matrixT}),
it is easy to see that $A_j$
is nonsingular.

\begin{prop} \label{prop:condition number}
The eigenvalues of $A_jA_j^\dag$ have the following form:
\be\label{expression of singular values}
(1-\gamma_j)^2 +2(1-\gamma_j)\cos\theta +1
=\left(\frac{\sin \theta}{\sin m\theta}\right)^2,
\ee
where $\theta$ is nonzero and  satisfies
\be\label{conditions of singular values}
(1-\gamma_j)\sin m\theta +  \sin(m+1)\theta=0.
\ee
\end{prop}

Before proving the above result, we first show how to estimate the condition number of $A$ from this proposition.

\begin{prop}
\label{thm:condition-special case}
Assuming that $d=1$, the condition number $\kappa$ of the linear system (\ref{linear system:forward method}) is
$\kappa=\Theta(m)$. Moreover,
$\|A\|=\Theta(1),\|A^{-1}\|=\Theta(m)$.
\end{prop}

\begin{proof}
Let $\sigma_{\max},\sigma_{\min}$ be the maximal and minimal nonzero singular value of $A$ respectively.
If $j=0$, then $\gamma_j=0$ and $A_j=\mathcal{T}$. The  singular values of $\mathcal{T}$ are
\be \label{singular values of T}
2  \cos \frac{k\pi}{2m+1} ,
\ee
where $k=1,\ldots,m$. A proof of this will be given at the end of this appendix. 
If we choose $k=m$, then
\be \label{upper bound:min}
\sigma_{\min} \leq 2  \cos \frac{m\pi}{2m+1} = 2 \sin \frac{\pi}{2(2m+1)} \leq \frac{\pi}{2m+1}.
\ee

To compute the minimal nonzero value of 
$(\sin \theta/\sin m\theta)^2$ in the interval
$[0,\pi]$, it suffices
to focus on the interval $\theta\in[0,\pi/2]$, since $|\sin m\theta|$ is periodic in the interval $[0,\pi/2]$, and the periods are
$\{[k\pi/m,(k+1)\pi/m]:k=0,\ldots,m/2-1\}$.
Also, in the interval $[0,\pi/2]$, $\sin \theta$
is increasing. Since we want to compute the
minimal value, we just need to consider
the interval $[0,\pi/m]$. Actually, we only
need to focus on $[0,\pi/2m]$ because
$|\sin m\theta|$ is symmetric along the line
$\theta=\pi/2m$. When $\theta$ is small, $\sin \theta\geq 2\theta/\pi$ and 
$\sin m\theta \leq m\theta$,
so
\be \label{lower bound:min}
\sigma_{\min} \geq \min_{0<\theta<\pi} \left|\frac{\sin \theta}{\sin m\theta}\right|
\geq  \frac{2}{m\pi} .
\ee
Therefore, we have
\be \label{min}
\sigma_{\min} = \Theta(1/m).
\ee

Next, we estimate $\sigma_{\max}$. 
Since $\alpha \Delta t/\Delta x^2 \leq 1/2$, we have
$0\leq \gamma_j\leq 2$. Thus,
$(1-\gamma_j)^2 +2(1-\gamma_j)\cos\theta +1
\leq 4$.
When $\gamma_j=1$, the eigenvalue is 1,
so $\sigma_{\max}\geq 1$. Note that
in the case $\alpha \Delta t/\Delta x^2=1/2$, then $\gamma_j=1$ implies that
$j=n/4$ in equation (\ref{gamma}).
As a result, $\sigma_{\max}=\Theta(1)$. 
Together with equation (\ref{min}), we obtain the claimed result.
\end{proof}

Next, we consider the general case $d>1$. It is easy to see that
\be
\label{eq:generall}
\mathcal{L} =  I_{n}^{\otimes d} +
\frac{\alpha \Delta t}{\Delta x^2} 
\sum_{j=1}^d 
I_{n}^{\otimes (j-1)} \otimes H \otimes I_{n}^{\otimes (d-j)}.
\ee
The coefficient matrix of the linear system (\ref{linear system:forward method}) is
\be
A = \mathcal{T} \otimes \mathcal{L}
- \frac{\alpha \Delta t}{\Delta x^2} 
\sum_{j=1}^d 
I_{n}^{\otimes (j-1)} \otimes H \otimes I_{n}^{\otimes (d-j)}.
\ee

\begin{repthm}{thm:condition}
The largest and smallest singular values of the matrix in (\ref{linear system:forward method}) satisfy $\sigma_{\max} = \Theta(1)$, $\sigma_{\min} = \Theta(1/m)$, respectively. Hence the condition number is $\Theta(m)$.
\end{repthm}

\begin{proof}
The proof of this theorem is similar to that of Proposition \ref{thm:condition-special case}. The calculation of the singular values of $A$ can be reduced to calculating the singular values of
\be
A_{j_1,\ldots,j_d} = (1-\gamma_{j_1}-\cdots-\gamma_{j_d}) \mathcal{T}
+(\gamma_{j_1}+\cdots+\gamma_{j_d}) I,
\ee
where $j_1,\ldots,j_d \in\{0,1,\ldots,n-1\}$. The result of Proposition \ref{prop:condition number} also holds for $A_{j_1,\ldots,j_d} $ by changing $\gamma_j$ into $\gamma_{j_1}+\cdots+\gamma_{j_d}$. Let $\sigma_{\max},\sigma_{\min}$ be the maximal and minimal nonzero singular value respectively. 

The estimation of $\sigma_{\min}$ is the same as that in the proof of Proposition \ref{thm:condition-special case}.
The upper bound is obtained by considering the special case $\gamma_{j_1}=\cdots=\gamma_{j_d}=0$.
Similarly to equation (\ref{upper bound:min}), $\sigma_{\min}\leq \pi/(2m+1)$. As for the lower bound, 
the proof of that in equation (\ref{lower bound:min})  is independent of $\gamma_j$, so it is also true 
for $A_{j_1,\ldots,j_d}$.
Thus $\sigma_{\min}=\Theta(1/m)$. 

As for $\sigma_{\max}$, if we consider the special case $\gamma_{j_1}=\cdots
=\gamma_{j_d}=1/d$,
then we obtain $\sigma_{\max} \geq 1$.
This special case is obtained by taking $j=n/4$ in the case $d\alpha \Delta t/\Delta x^2 = 1/2$. 
Since the eigenvalue of $A_{j_1,\ldots,j_d}$ also has the form
(\ref{expression of singular values}) by 
changing $\gamma_j$ into $\gamma_{j_1}+\cdots+\gamma_{j_d}$, $\gamma_j = 4 \frac{\alpha \Delta t}{\Delta x^2} \sin^2 \frac{j \pi}{n}$
and $d\alpha \Delta t/\Delta x^2 \leq  1/2$, we have
$\gamma_{j_1}+\cdots+\gamma_{j_d} \leq 
4 d\alpha \Delta t/\Delta x^2 \leq 2.$
By equation (\ref{expression of singular values}), $\sigma_{\max} \leq 4$.
Thus $\sigma_{\max}=\Theta(1)$,
and $\sigma_{\min}=\Theta(1/m)$.
\end{proof}

\begin{proof}[Proof of Proposition \ref{prop:condition number}]
For convenience, set $\beta_j=\gamma_j/(1-\gamma_j)$, then

\bea \label{sub-matrix}
A_jA_j^\dag &=& (1-\gamma_j)^2 \left(
  \begin{array}{cccccc}
    (1+\beta_j)^2 & -(1+\beta_j)  &        &&                  \\
    -(1+\beta_j)  & 1+(1+\beta_j)^2 & -(1+\beta_j)              \\
       & -(1+\beta_j)  & \ddots  & \ddots         \\
       &    & \ddots  & \ddots  & -(1+\beta_j)       \\
       &    &    &  -(1+\beta_j) & 1+(1+\beta_j)^2
  \end{array}
\right)  \\
&=& (1-\gamma_j)^2 [ (1+(1+\beta_j)^2) I_m - (1+\beta_j) Q_j],
\eea
where
\be
Q_j = \left(
  \begin{array}{cccccc}
    q_j & 1  &        &&                  \\
    1  & 0 & 1                        \\
       & 1  & \ddots  & \ddots         \\
       &    & \ddots  & \ddots  & 1       \\
       &    &    &  1 & 0
  \end{array}
\right),
\ee
and $q_j=1/(1+\beta_j)=1-\gamma_j$. 
In the following,
we need to compute the eigenvalues of
$Q_j$. The following lemma describes
the characteristic polynomial of $Q_j$.
It is easy to calculate that
$\det(Q_j+2I)=m+1+mq_j\neq 0$
as $-1\leq q_j\leq 1$.
This means $-2$ is not an eigenvalue of $Q_j$. In the following analysis,
we will not consider this case.

\begin{lem} 
\label{lem:char-poly}
Assume that $\lambda \neq 2$.
For any $m\geq 1$, let
\be
f_m=\left|
  \begin{array}{cccccc}
    \lambda & 1  &        &&                  \\
    1  & \lambda & 1                        \\
       & 1  & \ddots  & \ddots         \\
       &    & \ddots  & \ddots  & 1       \\
       &    &    &  1 & \lambda
  \end{array}
\right|_{m\times m}.
\ee
Then
\be
f_m = \frac{x_1^{m+1} - x_2^{m+1}}{x_1-x_2},
\ee
where $x_1 = \frac{1}{2} (\lambda + \sqrt{\lambda^2-4}), x_2 = \frac{1}{2} (\lambda - \sqrt{\lambda^2-4})$, and $x_1\neq x_2$.
Moreover,
\be \label{lem:char-poly-eq}
|Q_j+\lambda I| = q_j \frac{x_1^{m} - x_2^{m}}{x_1-x_2} 
+ \frac{x_1^{m+1} - x_2^{m+1}}{x_1-x_2}.
\ee
\end{lem}

\begin{proof}

By definition, $f_m = \lambda f_{m-1} - f_{m-2}$, then $f_m = \alpha_1 x_1^m + \alpha_2 x_2^m$ for some $\alpha_1,\alpha_2$.
Since $f_1=\lambda, f_2 = \lambda^2-1$, we have
\bea
\alpha_1 x_1 + \alpha_2 x_2     &=& \lambda, \\
\alpha_1 x_1^2 + \alpha_2 x_2^2 &=& \lambda^2-1.
\eea
Solving the linear system gives
\be
\alpha_1 = \frac{x_1}{x_1-x_2}, \alpha_2 = \frac{x_2}{x_2-x_1}.
\ee
So $f_m = \frac{x_1^{m+1} - x_2^{m+1}}{x_1-x_2}$. 
Since $\lambda \neq 2$, we obtain $x_1\neq x_2$.
By definition, 
\be
|Q_j+\lambda I| = (q_j+\lambda) f_{m-1} - f_{m-2} = q_j f_{m-1}+f_m
=q_j \frac{x_1^{m} - x_2^{m}}{x_1-x_2} + \frac{x_1^{m+1} - x_2^{m+1}}{x_1-x_2}.
\ee
This completes the proof.
\end{proof}

Now we have to solve for $\lambda$ from
equation (\ref{lem:char-poly-eq}), i.e.,
\be \label{eigen-equation0}
q_j (x_1^{m} - x_2^{m}) + (x_1^{m+1} - x_2^{m+1}) = 0.
\ee
Divides both sides of the above equation by $x_2^{m+1}$,
we obtain
\be
q_j \left(\frac{x_1^m}{x_2^m} - 1\right)\frac{1}{x_2} 
+ \left(\frac{x_1^{m+1}}{x_2^{m+1}} - 1\right) = 0.
\ee
Since $x_1x_2=1$, we have
\be \label{eigen-equation}
q_j(x_1^{2m}-1)x_1 + (x_1^{2(m+1)} - 1) = 0.
\ee
If $x_1$ is an solution, then $x_2=1/x_1$ is also a solution of the above 
equation. 
Assume that $x_1 = re^{i \theta}$.
Since $x_1+x_1^{-1} = \lambda \in \mathbb{R}$, if $\theta\neq 0 \mod \pi$, then $r=\pm 1$.

By (\ref{sub-matrix}) and noting that in Lemma \ref{lem:char-poly}, $-\lambda$
is the eigenvalue of $Q_j$, we obtain that
the eigenvalues of $A_jA_j^\dag$ are of the form
\bea
\sigma &=& (1-\gamma_j)^2[ 1+(1+\beta_j)^2 + (1+\beta_j) \lambda] \\
&=& (1-\gamma_j)^2[1+(1+\beta_j)^2 + (1+\beta_j) (x_1+x_1^{-1}) ] \\
&=& (1-\gamma_j)^2[ (1+x_1(1+\beta_j))(1+\frac{1+\beta_j}{x_1})],
\eea
where $x_1$ runs over all solutions of equation (\ref{eigen-equation0}).
By equation (\ref{eigen-equation}) and $q_j=1/(1+\beta_j)$, 
we know that $x_1^{2m+1} (1+x_1 (1+\beta_j)) = x_1+(1+\beta_j)$.
Thus $\sigma/(1-\gamma_j)^2$ can be rewritten as
\be \label{expression of eigenvalues}
x_1^{2m}(1+x_1 (1+\beta_j))^2 ~{\rm or}~
\frac{1}{x_1^{2m}}\left(1+\frac{1+\beta_j}{x_1}\right)^2.
\ee
If $x_1 \in \mathbb{R}$, and
if $|x_1|\geq 1$, then the 
first expression of (\ref{expression of eigenvalues}) implies that 
$\sigma/(1-\gamma_j)^2$ is 
exponentially large; however the second
expression shows that $\sigma/(1-\gamma_j)^2$ tends to zero.
The same contradiction also appears
if $|x_1|\leq 1$. So if $x_1 \in \mathbb{R}$, then $x_1=\pm 1$.
We prove this more formally in the following lemma.

\begin{lem}
If $x_1\in\mathbb{R}, |x_1| \geq 1$ 
and $x_1^{2m+1} (1+x_1(1+\beta_j)) = x_1+1+\beta_j$, then $x_1=\pm1$.
\end{lem}

\begin{proof}
First assume $x_1>1$. We have $x_1^{2m} (1+x_1 (1+\beta_j)) = 1+\frac{1+\beta_j}{x_1}$. 
The left side is strictly greater than $1+(1+\beta_j)$, 
while the right side strictly smaller than $1+(1+\beta_j)$, a contradiction.
Next assume $x_1<-1$. Set $\tilde{x}_1=-x_1>1$, then we have 
$(1+\beta_j)-\tilde{x}_1=\tilde{x}_1^{2m+1} (\tilde{x}_1(1+\beta_j)-1) 
\geq \tilde{x}_1(1+\beta_j)-1 > (1+\beta_j)-1$. This means $\tilde{x}_1<1$, a contradiction.
\end{proof}

Due to the two equivalent expressions (\ref{expression of eigenvalues}) of eigenvalues, it is also a contradiction if $0<|x_1|<1$. Since $x_1\neq x_2$, the above lemma
means $x_1 \notin \mathbb{R}$,
thus the only possibility is $x_1=e^{i\theta}$ for some $\theta$, 
then $(x_1^{2m}-1)x_1 + (1+\beta_j)(x_1^{2(m+1)} - 1) = 0$ implies that
\be
(x_1^{m}-x_1^{-m}) + (1+\beta_j)(x_1^{m+1} - x_1^{-m-1}) = 0.
\ee
So $(e^{i m\theta}-e^{-i m\theta}) + (1+\beta_j)(e^{i (m+1)\theta} 
- e^{-i (m+1)\theta}) = 0$, that is
\be \label{relation}
\sin m\theta + (1+\beta_j) \sin(m+1)\theta=0.
\ee
Thus,
\bea
\frac{\sigma}{(1-\gamma_j)^2} &=& (x_1^{m} (1+ x_1 (1+\beta_j)))^2 \\ 
&=& ((\cos m\theta+i \sin m\theta) (1+(1+\beta_j) \cos \theta 
+ i (1+\beta_j) \sin \theta))^2 \\ 
&=& [(\cos m\theta (1+(1+\beta_j) \cos \theta) 
- (1+\beta_j) \sin m\theta\sin \theta) \\
& & + \, i((1+\beta_j)\cos m\theta \sin\theta + 
(1+(1+\beta_j)\cos\theta)  \sin m\theta)]^2 \\
&=& [(\cos m\theta + (1+\beta_j) \cos(m+1)\theta)
+i(\sin m\theta + (1+\beta_j) \sin(m+1)\theta)]^2 \\
&=& (\cos m\theta + (1+\beta_j) \cos(m+1)\theta)^2  \\
&=&  \left(\frac{\sin \theta}{\sin(m+1)\theta}\right)^2,
\label{minimal singular value}
\eea
where the last identity (\ref{minimal singular value}) is derived from the identity (\ref{relation}).

On the other hand, 
\be \label{relation2}
\frac{\sigma}{(1-\gamma_j)^2} = 1+(1+\beta_j)^2 + (1+\beta_j)(x_1+x_1^{-1}) 
= 1+(1+\beta_j)^2 +2(1+\beta_j)\cos\theta.
\ee
Substitute $\beta_j = \gamma_j/(1-\gamma_j)$ 
into (\ref{relation}) and (\ref{relation2})
will yield the claimed results.
\end{proof}

Based on the above calculation, next we compute the singular values of $\mathcal{T}$, which is claimed in equation (\ref{singular values of T}). It suffices to choose $j=0$ in (\ref{expression of eigenvalues}). If $j=0$, then $\gamma_j=\beta_j=0$, so $x_1$ satisfies 
$x_1^{2m+1}(1+x_1) = (1+x_1)$. 
Since $x_1\neq -1$,
we obtain
$x_1^{2m+1}=1$, i.e., $e^{i(2m+1)\theta}=1$, thus $\theta = \frac{2k\pi}{2m+1},$ where $k=0,\pm1,\ldots,\pm m$. 
Note that $x_1 \neq x_2$, so $k\neq 0$.
Also note that $x_1x_2=1$, so we just need to
choose $k=1,2,\ldots,m$ to determine $x_1$.
For these $\theta$,
\be
\sigma = \left(\frac{\sin \frac{2k\pi}{2m+1}}{\sin\frac{2k(m+1)\pi}{2m+1}}\right)^2
=  \left(\frac{2 \sin \frac{k\pi}{2m+1} \cos \frac{k\pi}{2m+1} }{\sin\frac{k\pi}{2m+1}}\right)^2
=  \left( 2  \cos \frac{k\pi}{2m+1}  \right)^2.
\ee
Therefore, the singular values of $\mathcal{T}$
are $2  \cos \frac{k\pi}{2m+1}$, where $k=1,2,\ldots,m$.

%-------------------------------------------------------------------------------

\section{$\mathcal{L}$ is well-conditioned on nonnegative vectors}
\label{app:conditiononpositive}

In this appendix, we show that $\mathcal{L}$ cannot shrink nonnegative vectors too much, implying that the quantum algorithm for solving linear equations can construct a quantum state corresponding to $\mathcal{L}\mathbf{u_0}$ efficiently, given a quantum state corresponding to $\mathbf{u_0}$.

\begin{replem}{lem:conditiononpositive}
Let $\mathcal{L}$ be defined by (\ref{eq:linstep}), taking $\Delta t = \Delta x^2/(2\alpha d)$ as in Corollary \ref{cor:accuracy}. Then for all nonnegative vectors $\mathbf{u}$, $\|\mathcal{L}\mathbf{u}\|_2^2 / \|\mathbf{u}\|_2^2 \ge 1/(2d)$.
\end{replem}

\begin{proof}
Write $\mathcal{L} = \sum_{i=1}^d \mathcal{L}_i$, where $\mathcal{L}_i$ acts only on the $i$'th coordinate and
\be \mathcal{L}_i \widetilde{u}(x,t) = \frac{1}{2d} \left( \widetilde{u}(\dots,x_i+\Delta x,\dots,t) + \widetilde{u}(\dots,x_i-\Delta x,\dots,t) \right). \ee
This operator corresponds to the matrix
\be \frac{1}{d} \begin{pmatrix} 0 & \frac{1}{2} & & \dots & \frac{1}{2}\\
\frac{1}{2} & 0 & \frac{1}{2} & \dots & \\
 & \frac{1}{2} & 0 & \ddots & \\
 & & \ddots & & \\
\frac{1}{2} & 0 & \dots & \frac{1}{2} & 0
\end{pmatrix}. \ee
Then
\be \|\mathcal{L}\mathbf{u}\|_2^2 = \sum_{i,j=1}^d \mathbf{u}^T \mathcal{L}_i \mathcal{L}_j \mathbf{u} \ge \sum_{i=1}^d \mathbf{u}^T \mathcal{L}_i^2 \mathbf{u} \ee
using non-negativity of $\mathcal{L}_i$ and $\mathbf{u}$. It is easy to see that the matrix for $\mathcal{L}_i^2$ has entries all equal to $\frac{1}{2d^2}$ on the main diagonal, and non-negative entries elsewhere. Therefore, for each $i$,
\be \mathbf{u}^T \mathcal{L}_i^2 \mathbf{u} \ge \frac{\|\mathbf{u}\|_2^2}{2d^2}, \ee
and hence $\|\mathcal{L}\mathbf{u}\|_2^2 \ge \|\mathbf{u}\|_2^2/(2d)$.
\end{proof}

%-------------------------------------------------------------------------------

\section{Bounds on $\ell_2$ norm of solutions to heat equation}
\label{app:l2bound}

In this appendix we prove Lemma \ref{lem:l2bound}, which gives upper and lower bounds on $\|\mathcal{L}^\tau \ket{0}\|_2^2$ in the special case where $\Delta t = \Delta x^2/(2d\alpha)$. To achieve this, we will use Fourier analysis (similarly to Appendix \ref{appendix:Estimation of the condition number}). As in the previous appendix, write $\mathcal{L} = \sum_{i=1}^d \mathcal{L}_i$, where $\mathcal{L}_i$ acts only on the $i$'th coordinate and
\be \mathcal{L}_i \widetilde{u}(x,t) = \frac{1}{2d} \left( \widetilde{u}(\dots,x_i+\Delta x,\dots,t) + \widetilde{u}(\dots,x_i-\Delta x,\dots,t) \right). \ee
Each operator $\mathcal{L}_i$ is diagonalised by the quantum Fourier transform on $\Z_n$ and has eigenvalues $\frac{1}{d} \cos(2 \pi y / n)$ for $y=0,\dots,n-1$. Applying the quantum Fourier transform to $\ket{0}$ gives a uniform superposition over all Fourier modes $y$, which we identify with elements of $\Z_n$. Then
\be \label{eq:ltl2norm} \|\mathcal{L}^\tau \ket{0}\|_2^2 = n^{-d} \sum_{y_1,\dots,y_d=0}^{n-1} \left[\frac{1}{d} \sum_{i=1}^d \cos(2\pi y_i / n) \right]^{2\tau}. \ee
We also observe that $\mathcal{L}$ describes a simple random walk on a periodic $d$-dimensional square lattice. As
\be \|\mathcal{L}^\tau \ket{0}\|_2^2 = \braket{0|\mathcal{L}^{2\tau}|0}, \ee
where we use $\ket{0}$ to denote the origin, we can interpret $\|\mathcal{L}^\tau \ket{0}\|_2^2$ as the probability of returning to the origin after $2\tau$ steps of the random walk.

To complete the proof of Lemma \ref{lem:l2bound} and bound this quantity, we will first handle the simpler 1-dimensional case separately.

\begin{lem}
\label{lem:1dwalk}
Let $d=1$ and let $\mathcal{L}$ be defined by (\ref{eq:linstep}), taking $\Delta t = \Delta x^2/(2\alpha)$ as in Corollary \ref{cor:accuracy}. Then
\be \max\left\{\frac{1}{n}, \frac{1}{2\sqrt{\tau}} \right\} \le \braket{0|\mathcal{L}^{2\tau}|0} \le \frac{4}{n} + \frac{1}{\sqrt{\pi \tau}}. \ee
\end{lem}

\begin{proof}
%The case where $k$ is odd is immediate, so we henceforth set $k=2t$ for integer $t$, and begin with the lower bounds.
A lower bound
\be \braket{0|\mathcal{L}^{2\tau}|0} \ge \frac{\binom{2\tau}{\tau}}{2^{2\tau}} \ge \frac{1}{2\sqrt{\tau}} \ee
follows by observing that the probability of returning to 0 after $2\tau$ steps is lower-bounded by the probability of a random walk on the integers (not considered modulo $n$) returning to 0 after $2\tau$ steps, which is exactly $\binom{2\tau}{\tau}/2^{2\tau}$. Next, we use (\ref{eq:ltl2norm}) to obtain
\be \braket{0|\mathcal{L}^{2\tau}|0} = \frac{1}{n} \sum_{y=0}^{n-1} \cos(2\pi y / n)^{2\tau}, \ee
which is an exact statement for the walk modulo $n$, and observe that a lower bound of $1/n$ is immediate from considering only the $y=0$ term.

For an upper bound, we start with the same expression, and use
\bea \braket{0|\mathcal{L}^{2\tau}|0} &\le& \frac{4}{n}\sum_{y=0}^{\lfloor n/4 \rfloor} \cos(2\pi y / n)^{2\tau}\\
&\le& \frac{4}{n} \sum_{y=0}^{\lfloor n/4 \rfloor} e^{-4\tau \pi^2 y^2 / n^2}\\
&\le& \frac{4}{n} \left(1 + \int_{0}^{\infty} e^{-(2\sqrt{\tau} \pi y/n)^2} dy \right) \\
&=& \frac{4}{n} \left(1 + \frac{n}{2 \pi \sqrt{\tau}} \int_{0}^{\infty} e^{-y^2} dy \right) \\
&=& \frac{4}{n} + \frac{1}{\sqrt{\pi \tau}}.
%&\le& n^d \sum_{y_1,\dots,y_d=0}^{n-1} \left[ \frac{1}{d}\sum_{i=1}^d \cos(2\pi y_i / n)^{2t} \right]\\
%&=& \frac{n^d}{d} \sum_{i=1}^d \sum_{y_1,\dots,y_d=0}^{n-1} \cos(2\pi y_i / n)^{2t} \\
%&=& \frac{n^d}{d} \sum_{i=1}^d \sum_{y_i=0}^{n-1} \cos(2\pi y_i / n)^{2t} \sum_{y_1,\dots,y_{i-1},y_{i+1},\dots,y_d=0}^{n-1} 1\\
%&=& n^{2d-1} \sum_{y_1=0}^{n-1} \cos(2\pi y_i / n)^{2t}.
\eea
The first inequality follows from splitting the sum up as
\be
\sum_{y=0}^{n-1} \cos(2\pi y / n)^{2\tau} = \!\!\!\!\sum_{y\le n/4} \cos(2\pi y / n)^{2\tau} + \!\!\!\!\!\!\!\! \sum_{n/4 < y\le n/2} \!\!\!\!\!\!\!\!\cos(2\pi y / n)^{2\tau} +\!\!\!\!\!\!\!\!\sum_{n/2 < y\le 3n/4} \!\!\!\!\!\!\!\!\cos(2\pi y / n)^{2\tau} +\!\!\!\!\!\!\!\! \sum_{3n/4 < y < n} \!\!\!\!\!\!\!\!\cos(2\pi y / n)^{2\tau}.\ee
Using that $\cos(\theta)^2 = \cos(k\pi \pm \theta)^2$ for $k \in \Z$, each of the last three sums is upper-bounded by the first one. For example,
\be \sum_{n/4 < y \le n/2}\!\!\!\!\!\!\!\! \cos(2\pi y / n)^{2\tau} = \!\!\!\!\!\!\!\!\sum_{n/4 < y \le n/2}\!\!\!\!\!\!\!\! \cos(2\pi (n/2-y) / n)^{2\tau} = \!\!\!\!\!\!\!\!\sum_{n/4 < n/2 - y' \le n/2}\!\!\!\!\!\!\!\! \cos(2\pi y'/ n)^{2\tau} = \!\!\!\!\sum_{0 \le y' < n/4}\!\!\!\! \cos(2\pi y'/ n)^{2\tau}; \ee
note that if $n$ is not a multiple of 2, $y' = n/2-y$ ranges over values of the form $i + 1/2$ for integer $i$. As $\cos \theta$ is decreasing in the range $0 \le \theta \le \pi$, replacing the sum with a sum over integers in the range $\{0,\dots,n/4\}$ could not make it smaller. The second inequality uses that $\cos \theta \le e^{-\theta^2/2}$ for $\theta \le \pi/2$ \cite[Chapter 3, Theorem 2]{diaconis88}.
\end{proof}

\begin{replem}{lem:l2bound}
%Let $\mathbf{u_0}$ be defined by $\mathbf{u_0}(0) = 1$, $\mathbf{u_0}(\mathbf{x}) = 0$ for $\mathbf{x} \in G$, $\mathbf{x} \neq 0^d$.
Let $\mathcal{L}$ be defined by (\ref{eq:linstep}), taking $\Delta t = \Delta x^2/(2d\alpha)$ as in Corollary \ref{cor:accuracy}. Then for any $\tau \ge 1$,
\be \max\left\{\frac{1}{n^d}, \frac{1}{(4\sqrt{\tau})^d} \right\} \le \braket{0|\mathcal{L}^{2\tau}|0} \le d e^{-\tau/(4d)} + \left(\frac{4}{n} + \sqrt{\frac{d}{\pi \tau}}\right)^d. \ee
\end{replem}

\begin{proof}
We start by proving the upper bound, which is based on the interpretation of $\braket{0|\mathcal{L}^{2\tau}|0}$ as the probability of returning to the origin after $2\tau$ steps of a random walk. Each step corresponds to choosing one of $d$ dimensions uniformly at random, then moving in one of two possible directions in that dimension. The walk returns to the origin after $2\tau$ steps if it has done so in every dimension. To understand the probability of this event, we use Lemma \ref{lem:1dwalk}.

Let $s \in \{1,\dots,d\}^{2\tau}$ denote the sequence of dimensions chosen by the walk, and let $N_i(s)$ denote the number of $i$'s in $s$. Let $p(N)$ denote the probability that a 1d walk returns to the origin after $N$ steps. Then
\be \braket{0|\mathcal{L}^{2\tau}|0} = d^{-2\tau} \sum_{s \in \{1,\dots,d\}^{2\tau}} p(N_1(s)) \dots p(N_d(s)) \ee
using independence of the random walks, conditioned on $s$. By Lemma \ref{lem:1dwalk}, we have
\be \braket{0|\mathcal{L}^{2\tau}|0} \le d^{-2\tau} \sum_{s \in \{1,\dots,d\}^{2\tau}} \left(\frac{4}{n} + \frac{1}{\sqrt{\pi N_1(s)}}\right) \dots \left(\frac{4}{n} + \frac{1}{\sqrt{\pi N_d(s)}}\right). \ee
By a Chernoff bound, for each $i \in \{1,\dots,d\}$,
\be \Pr_{s \in \{1,\dots,d\}^{2\tau}}\left[N_i(s) \le \frac{\E_s[N_i(s)]}{2}\right] = \Pr_{s \in \{1,\dots,d\}^{2\tau}}\left[N_i(s) \le \frac{\tau}{d}\right] \le e^{-\tau/(4d)}, \ee
so using a union bound over $i$,
\begin{eqnarray}
\braket{0|\mathcal{L}^{2\tau}|0} &\le& d e^{-\tau/(4d)} + d^{-2\tau} \sum_{\substack{s \in \{1,\dots,d\}^{2\tau}\\\forall i,N_i(s) > \tau/d}} \left(\frac{4}{n} + \frac{1}{\sqrt{\pi N_1(s)}}\right) \dots \left(\frac{4}{n} + \frac{1}{\sqrt{\pi N_d(s)}}\right)\\
&\le& d e^{-\tau/(4d)} + d^{-2\tau} \sum_{\substack{s \in \{1,\dots,d\}^{2\tau}\\\forall i,N_i(s) > \tau/d}} \left(\frac{4}{n} + \sqrt{\frac{d}{\pi \tau}}\right)^d\\
&\le& d e^{-\tau/(4d)} + \left(\frac{4}{n} + \sqrt{\frac{d}{\pi \tau}}\right)^d
\end{eqnarray}
as claimed. Next we prove the lower bound. Using
\be \braket{0|\mathcal{L}^{2\tau}|0} = n^{-d} \sum_{y_1,\dots,y_d=0}^{n-1} \left[ \frac{1}{d}\sum_{i=1}^d \cos(2\pi y_i / n) \right]^{2\tau}, \ee
we get a lower bound of $n^{-d}$ immediately by considering the term $y_1=\dots=y_d=0$.
For the remaining part of the lower bound, we use that from Lemma \ref{lem:1dwalk}, the probability that a walk on $\Z_n$ making $2k$ steps returns to the origin is lower-bounded by $\frac{1}{2\sqrt{k}}$. So, if each of the $d$ independent random walks makes an even number of steps, the probability that they all simultaneously return to the origin is at least $\frac{1}{(2\sqrt{\tau})^d}$. It remains to lower-bound the probability that all of the walks make an even number of steps.

Let $N_e(d,2\tau)$ denote the number of sequences of $2\tau$ integers between 1 and $d$ such that the number of times that each integer appears in the sequence is even. The probability that all the walks make an even number of steps is $N_e(d,2\tau) / d^{2\tau}$. We will show by induction on $d$ that $N_e(d,2\tau) \ge d^{2\tau} / 2^d$. For the base case, $N_e(1,2\tau) = 1 \ge 1/2$ as required. Then for $d \ge 2$,
\begin{eqnarray}
N_e(d,2\tau) &=& \sum_{i=0}^{\tau} \binom{2\tau}{2i} N_e(d-1,2\tau-2i)\\
&\ge& \sum_{i=0}^\tau \binom{2\tau}{2i} \frac{1}{2^{d-1}} (d-1)^{2\tau-2i}\\
&=& (d-1)^{2\tau} \frac{1}{2^{d-1}} \sum_{i=0}^\tau \binom{2\tau}{2i} (d-1)^{-2i}\\
&=& (d-1)^{2\tau} \frac{1}{2^{d-1}} \frac{1}{2}\left( \left(1+\frac{1}{d-1}\right)^{2\tau} + \left(1-\frac{1}{d-1}\right)^{2\tau}\right) \\
&=& \frac{1}{2^d} \left( d^{2\tau} + \left(d-2\right)^{2\tau}\right) \\
&\ge& \frac{1}{2^d} d^{2\tau}.
\end{eqnarray}
Therefore, with probability at least $1/2^d$, all of the walks make an even number of steps, and the probability that they all return to the origin after $2\tau$ steps in total is at least $\frac{1}{(4\sqrt{\tau})^d}$ as claimed.
\end{proof}

\end{document}